%% file: msc_paper_arxiv.tex
\documentclass[aps,pre,reprint,onecolumn,notitlepage,groupedaddress]{revtex4-1}
\usepackage{epsfig,graphics,graphicx,amssymb,amsmath,color,mathtools,subeqnarray,bm,mathtools,upgreek}
\usepackage[colorlinks=true,linkcolor=blue]{hyperref}
\usepackage[sfdefault=cmbr]{isomath}

\usepackage{msc_notations}

\def \upi{\pi}
\def \mathsfbi{\mathsf{\bm}}
\def \boldsymbol{\bm}
\def \Rey{Re}
\def \bnabla{\bm{\nabla}}

\def \fullsquare{\mbox{$\blacksquare$}}
\def \fullcirc{\mbox{{\Large$\bullet$}}}
\def \fulltriangledown{\mbox{$\blacktriangledown$}}

\graphicspath{
{1_introduction/figures/}
{3_origin/figures/}
{4_advection/figures/}
{5_capture/figures/}
}

\begin{document}

\title{Vortex-induced vibrations: a soft coral feeding strategy?\footnote{Accepted for publication in the Journal of Fluid Mechanics, available at https://doi.org/10.1017/jfm.2021.252.}}
\author{Mouad Boudina}\email{Electronic mail: mouad.boudina@polymtl.ca}
\affiliation{Laboratory for Multiscale Mechanics (LM2), Polytechnique Montr\'{e}al, Montr\'{e}al, Qu\'{e}bec, Canada}
\affiliation{
Department of Mechanical Engineering, Polytechnique Montr\'{e}al, Montr\'{e}al, Qu\'{e}bec, Canada}

\author{Fr\'{e}d\'{e}rick P. Gosselin}
\affiliation{Laboratory for Multiscale Mechanics (LM2), Polytechnique Montr\'{e}al, Montr\'{e}al, Qu\'{e}bec, Canada}
\affiliation{
Department of Mechanical Engineering, Polytechnique Montr\'{e}al, Montr\'{e}al, Qu\'{e}bec, Canada}

\author{St\'{e}phane \'{E}tienne}
\affiliation{
Department of Mechanical Engineering, Polytechnique Montr\'{e}al, Montr\'{e}al, Qu\'{e}bec, Canada}


\begin{abstract}
Soft corals, such as the bipinnate sea plume \emph{Antillogorgia bipinnata}, are colony building animals that feed by catching food particles brought by currents. Because of their flexible skeleton, they bend and sway back and forth with the wave swell. In addition to this low-frequency sway of the whole colony, branches of \emph{A.~bipinnata} vibrate at high frequency with small amplitude and transverse to the flow as the wave flow speed peaks.
In this paper, we investigate the origin of these yet unexplained vibrations and consider their effect on soft corals. 
Estimation of dynamical variables along with finite element implementation of the wake-oscillator model favour vortex-induced vibrations (VIVs) as the most probable origin of the observed rapid dynamics. To assess the impact of the dynamics on filter feeding, we simulated particles advected by the flow around a circular cylinder and calculated the capture rate with an in-house monolithic fluid-structure interaction (FSI) finite element solver and Python code. We observe that vibrating cylinders can capture up to 40\% more particles than fixed ones at frequency lock-in. Therefore, VIVs plausibly offer soft corals a better food capture.
\end{abstract}

\maketitle

\input{1_introduction/introduction}
\input{3_origin/origin}
\input{4_advection/advection}
\input{5_capture/capture}
\input{6_conclusion/conclusion}

\bibliography{msc_biblio}

\end{document}

%% file: 1_introduction/introduction.tex
\section{Introduction}

\begin{figure}
\centering
\includegraphics[width=\textwidth]{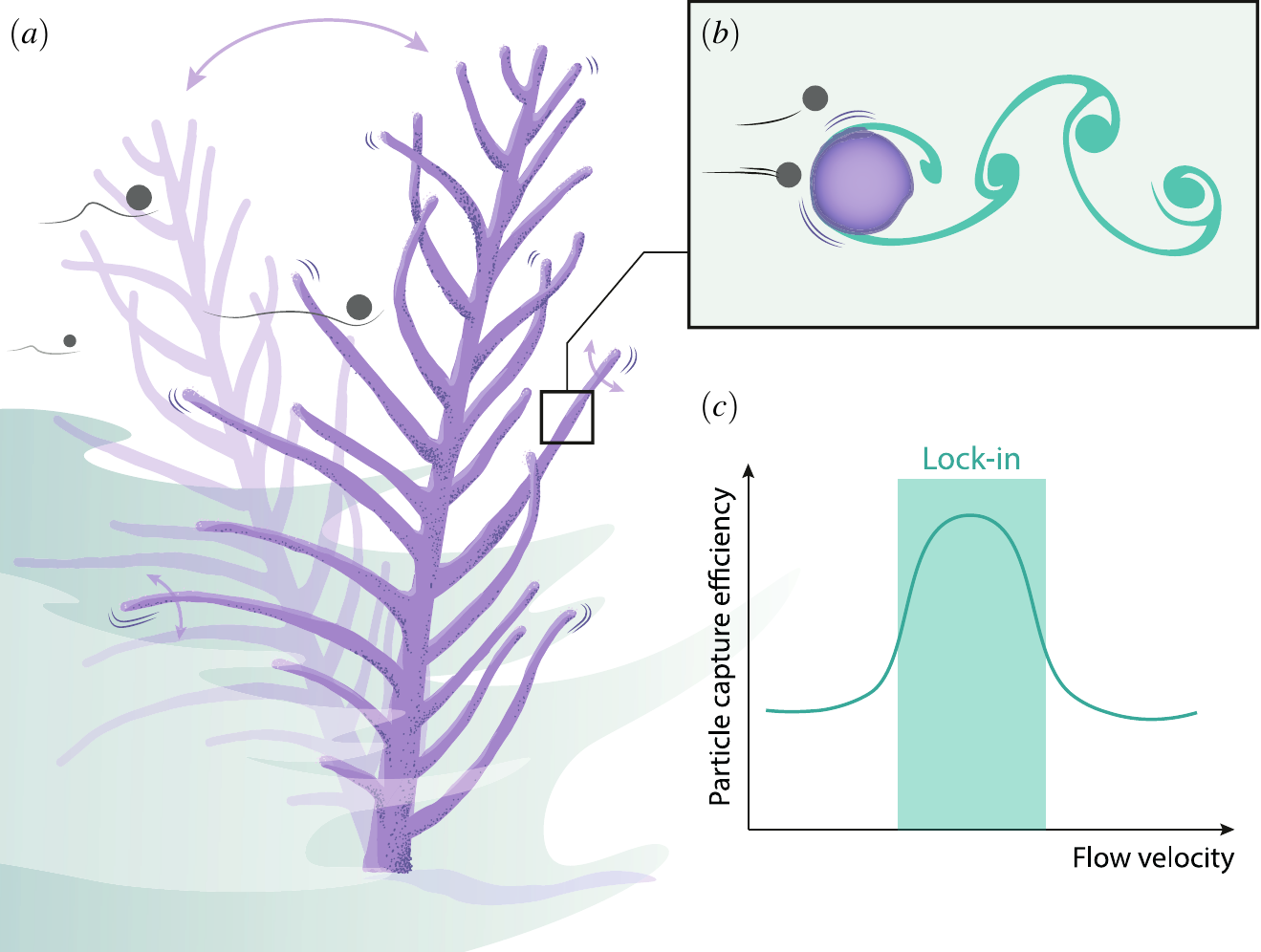}
\caption{($a$) Schematic of a bipinnate sea plume \textit{A. bipinnata} swaying back and forth under surface wave action, with branches exhibiting small and rapid transverse vibrations. The coral colony stands perpendicular to the flow and encounters incoming food particles. ($b$) Cross-section of a branch vibrating owing to the vortex shedding in the wake, and facing two incoming particles. The upper particle crosses over the branch and escapes capture, whereas the lower particle approaches the branch surface and gets intercepted. ($c$) Schematic summary of the numerical results of the particle capture efficiency, as defined in equation~(\ref{eq:original_def_eta}), versus flow speed. The soft coral branch maximises the efficiency of particle capture during the lock-in phase, where the vortex shedding frequency synchronises with the natural frequency of the branch. Art by IMPAKT Scientifik.}
\label{fig:figure1}
\end{figure}

Corals thrive in nutrient-poor waters, where seaweeds and seagrasses do not grow. Corals can do this, in part, because they benefit from the photosynthesis of their symbiotic algae zooxanthellae, but also by passively filter-feeding on food particles brought by the ambient water flow \citep{ribes_heterotrophic_1998}. 
Soft corals adopt various strategies to intercept food particles. For example, the sea fans \emph{Gorgonia ventalina} and \emph{Gorgonia flabellum} grow with their skeleton plane perpendicular to the predominant current, hence maximising their surface area normal to the flow \citep{wainwright_orientation_1969}. 
The bipinnate sea plume \emph{Antillogorgia bipinnata}, in the same  \emph{Gorgoniidae} family as sea fans \citep{williams_resurrection_2012}, grows to form flexible colonies with an arborescent morphology as schematised in figure \ref{fig:figure1}($a$).
\citet{gosselin_mechanics_2019} describes a peculiar motion of \emph{A.~bipinnata}: it sways back and forth under the low frequency ($\sim 0.4$ Hz) forcing of wave action. When the water flow speed peaks in the surface wave cycle, the branches of \textit{A.~bipinnata} undergo high frequency ($\sim 7$ Hz) vibrations transverse to the flow. 
In the present paper, we show how vortex-induced vibrations are responsible for the high frequency vibration of \emph{A.~bipinnata}, and how the dynamics increase the food particle capture efficiency, as shown in figure \ref{fig:figure1}.

To cope with the ambient fluid flows, arborescent species, on land or underwater, rely chiefly on their flexibility. Under a steady current, they are known to streamline and reconfigure their shape to minimise the hydrodynamic drag and prevent breakage \citep{vogel_drag_1984, de_langre_effects_2008, gosselin_mechanics_2019}. This incentive is amongst the reasons why soft corals, along with other species, modulate the length and stiffness of their branches \citep{jeyasuria_mechanical_1987, sanchez_phenotypic_2007}.
Under oscillatory flow, such as wave surge, their stems sway back and forth periodically and exhibit specific deformation profiles depending on the wave frequency and water speed \citep{leclercq_reconfiguration_2018, lei_blade_2019}.
As for the interaction of the flow with structures having a branching pattern, the response is a complex motion involving a series of vibrational modes with frequencies close to each other \citep{rodriguez_scaling_2008, rodriguez_multimodal_2012, der_loughian_measuring_2014}. 
\textit{A.~bipinnata} exhibits all these behaviours: its main stem reconfigures and sways with the wave surges, and its branches vibrate rapidly  spanning  small transverse amplitudes.

Vortex-induced vibrations are common in a marine environment, which lead to small and rapid oscillations \citep{williamson_vortex-induced_2004, sarpkaya_wave_2010, fredsoe_hydrodynamics_2006}. In the case of a circular spring-mounted cylinder, they can appear for Reynolds numbers as low as 20 \citep{etienne_low_2012}, and become pronounced when the von K\'{a}rm\'{a}n vortex street establishes in the wake. As the frequency of vortex shedding gets close to the natural frequency of the spring-mounted cylinder, the structure and the wake dynamics synchronise and amplify. This state is known as frequency synchronisation or \textit{lock-in} \citep{bishop_lift_1964}. Flexible structures may exhibit several frequency lock-in ranges owing to the multiplicity of their natural frequencies \citep{chaplin_laboratory_2005}. The VIVs along flexible structures propagate either as travelling waves when the structure has free endpoints \citep{newman_direct_1997, facchinetti_vortex-induced_2004}, standing waves when it has pinned endpoints \citep{evangelinos_dynamics_1999}, or a mix of both modes when the flow is non-uniform \citep{lucor_riser_2006, violette_computation_2007}.
In the living world, vortices have been credited as a potential factor in biological processes such as spore dispersal on plants \citep{kim_vortex-induced_2019} and particle feeding of starfish larvae \citep{gilpin2017}. These larvae spend significant energy to seed vortices around their body to the detriment of their swimming speed, but, in return, they expand their filtering area and capture more food particles. It is also suggested qualitatively that black fly and mosquito larvae, being suspended in water, create vortices as a strategy to drive more food up to their fans \citep{chance_hydrodynamics_1986, widahl_flow_1992}. By increasing the food particle capture rate, we believe that VIVs might be a boon to soft corals.

Soft corals are colonies of polyps, which are small tubular organisms with tentacles, a mouth, and a digestive system \citep{fabricius_octocorallia_2011, veron_corals_2011}. They are passive filter feeders \citep{ribes_heterotrophic_1998}, waiting passively for water currents to carry its particulate nutritional content (e.g. detritus, organic debris, phytoplankton, protists). When an edible particle comes close to a polyp, the tentacles stretch and catch the particle (encounter phase), then retain it by activating spines or secreting mucus (retention phase) \citep{shimeta_mechanisms_1997}. 
The study of particle capture is important to evaluate the filtering efficiency of such species and connect it with the flow properties. In the particle filtering literature, especially in the field of fluid mechanics and chemistry, encounter and retention are combined into a single phase termed `capture' or `interception'.
Multiple studies have considered the particle capture of a fixed circular cylinder \citep{weber_interceptional_1983, palmer_observations_2004, haugen_particle_2010, espinosa-gayosso_particle_2012, espinosa-gayosso_particle_2013}.
\citet{weber_interceptional_1983} and \citet{palmer_observations_2004} defined the capture efficiency as
\begin{equation}
\eta = \frac{\Ndot}{\Ndot\subrm{init}},
\label{eq:original_def_eta}
\end{equation}
where $\Ndot$ is the rate of captured particles, and $\Ndot\subrm{init}$ is the rate of particles released from an opening the same size as the cylinder.
Theoretical \citep{weber_interceptional_1983}, experimental \citep{palmer_observations_2004}, and numerical \citep{haugen_particle_2010, espinosa-gayosso_particle_2012, espinosa-gayosso_particle_2013} results agree that the capture efficiency increases with the cylinder-based Reynolds number and the particle size. The experiments of \citet{palmer_observations_2004} show, in addition, that cylinders with a rough surface capture particles more efficiently than those with a smooth surface. Surprisingly, data on particle capture by vibrating collectors are scarce. Even the handful of papers that have examined the effect of collector motion \citep{krick_adding_2015, mccombe_collector_2018} were imposing oscillation frequencies and amplitudes, not always covering the vibration parameter range encountered in the living world. Field experiments on a timothy grass revealed that it captured and germinated more pollen than transverse-tethered and stream-wise-tethered grasses \citep{mccombe_collector_2018}. This example indicates that it is important to allow the collector to be free to move instead of imposing oscillations. Not only is the motion more realistic, but the vibration amplitude and frequency vary simultaneously with the flow and cylinder parameters, and so cannot be decoupled. Although the idea that vibrations could improve particle capture was already suspected \citep{niklas_biophysical_2015}, there are no explicit mention of VIVs in the particle interception literature, and the variation of the capture efficiency with the relevant dynamical parameters remains unclear owing to the absence of quantitative data.

Understanding more about corals is critical as they form the basis of ecosystems that are home to countless fish and invertebrates. These coral reef ecosystems are under the multi-pronged threat of global warming: ocean acidification, water temperature rise and more frequent hurricane passage. All coral species are not affected to the same degree by these threats, as soft coral species seem to fare better in these changing environmental conditions \citep{inoue_spatial_2013,tsounis_three_2017}. The existing hydrodynamic studies of corals are mostly limited to the hard reef-building kind \citep{monismith_hydrodynamics_2007}. The biological role of soft coral flexibility in their fluid-structure interaction remains relatively unexplored.

This paper intends to fill this gap and prove that VIVs can increase the capture rate of a cylindrical branch. For this aim, we divide the ensuing work into two parts. We first present qualitative and quantitative arguments that put forward VIVs as the most probable source of the fast motion of soft coral branches. In the second part, we model the branch as a circular spring-mounted cylinder that is free to oscillate in the transverse and stream-wise directions, and compute its interaction with fluid flow using two-dimensional direct numerical simulations. We integrate the trajectories of particles advected by the flow, calculate the rate of capture, and compare the latter with the case of a fixed cylinder. Finally, using the results of the capture rate, we propose a link between the morphology of soft corals and the predominant local water speeds.

%% file: 3_origin/origin.tex
\section{Origin of vibrations}
\label{sec:origin}
\citet{gosselin_mechanics_2019} reported a vibrating bipinnate sea plume \textit{A.~bipinnata} on a SCUBA dive off of Isla Mujeres near Cancun, Mexico on March 25th 2015, at approximately 10m depth. The vibrational dynamics was similar to that seen in an online recorded video \citep{youtube_caribbean_2013}. Using the software \textsc{imageJ} \citep{schindelin_fiji_2012}, we assessed from that video that the flow velocity was $U_{0} \sim 10$~cm/s and the wave period was $T\subrm{wave} \sim 5$ s. With a branch diameter of $D \sim 2$~mm \citep{bayer_shallow-water_1961, cairns_checklist_1977}, and given the water kinematic viscosity was $\nuf \sim 10^{-6}$ m$^{2}$/s, the Reynolds number and the Keulegan-Carpenter number in the stream-wise direction were $\Rey = U_{0}D/\nuf \sim 200$ and $KC = U_{0}T\subrm{wave}/D \sim 250$, respectively, which meant that the flow was under a vortex shedding regime \citep{fredsoe_hydrodynamics_2006}. Finally, from an estimated value of the vibration frequency of $\fn \sim 7-9$ Hz, the reduced velocity was $\Ur = U_{0}/\fn D \sim 5-7$, which suggested VIVs lock-in dynamics.

As far as VIVs might be in play, we sought to reproduce the observed coral branch motion by simulating its reconfiguration and vibration under flow.
For this, we used \citet{mou3adb_rodics_2020}, a finite element solver of Kirchhoff rods based on the FEniCS platform \citep{alnaes_fenics_2015}. The dynamics of a branch was modelled with rod elements which considered the dynamical three-dimensional bending and torsion for an arbitrarily large deformation. This deformation arose from a static drag force based on the semi-empirical formulation of \cite{taylor_analysis_1952} and a dynamical coupling with the shed vortices, which were accounted in the wake-oscillator model of \citet{facchinetti_coupling_2004}.
Details on the Kirchhoff equations, the wake-oscillator model, as well as the verification and validation of \citet{mou3adb_rodics_2020} are provided in the other supplementary information available at https://doi.org/10.1017/jfm.2021.252.
We take again a representative diameter of $D \sim 2$~mm. We focused on branches having lengths of $L \sim$ 1, 5 and 8~cm.
Moreover, we considered a coral density equal to the fluid density $\rhos \sim \rhof \approx 10^{3}$ kg/m$^{3}$.
This choice is realistic because soft coral branches are not floating up or falling down, so they can be considered as approximately neutrally-buoyant.
As for the flexural rigidity, we conducted three-point bending tests on wet soft coral branches freshly taken out from an aquarium. We found a flexural rigidity of $EI \sim 5 \times 10^{-7}$ N.m$^{2}$.
We simulated a water flow of speed $U_{0} \sim 10$~cm/s, as estimated above from the online video.

\begin{figure}
\centerline{
\includegraphics{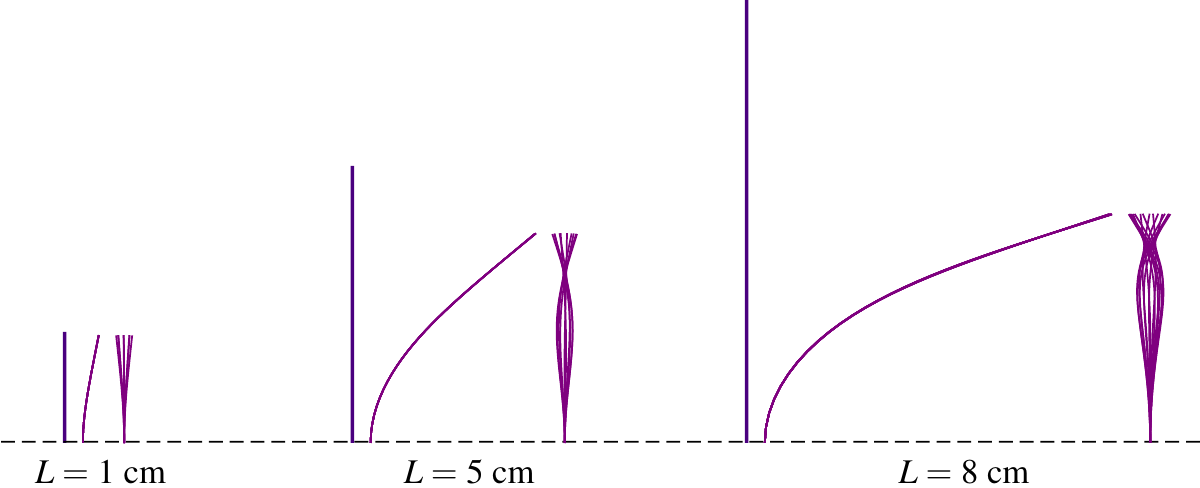}
}
\caption{Numerical deformation profiles of elastic rods of lengths $L = 1$, 5 and 8~cm subjected to VIV in a unidirectional flow. The vertical line in the left of each case indicates the initial configuration of the rod. On the middle and the right are the lateral and frontal profiles, respectively. The lengths are proportional. In the online record of the vibrating soft coral, we also see small branches staying relatively straight and having a first mode vibration, and longer branches reconfiguring more with the flow and having a second mode vibration with a single zero-displacement node.}
\label{fig:deformation_profiles}
\end{figure}

Numerical deformation profiles are shown in figure \ref{fig:deformation_profiles}. The first common feature in the simulations and real coral observations was the extent of bending in the water flow direction. Both the elastic rod and branches of $L \sim 1$~cm remained relatively straight, whereas those with $L \sim 5$ and 8~cm reconfigured more with the flow. The second common feature was the transverse envelope of vibrations. Short branches and rods exhibited a first modal shape with a maximum deflection at the tip, whereas longer branches had a second modal shape with a single zero-displacement node. These similarities between the simulations and the video observations were additional clues supporting our hypothesis that VIVs are the most probable origin of the observed rapid motion of soft coral branches.

Furthermore, we show that other types of flow-induced vibrations cannot be the principal cause of the sustained soft coral vibration. If the vibration was induced by an external excitation, such as turbulent buffeting, the branch frequency would match the peak frequency $\fpeak$ of the sea wave spectrum in the region.
From the National Data Buoy Center website \citep{national_data_buoy_center_national_nodate},
the peak frequencies of shallow waters in the Gulf of Mexico, where some soft corals live, range from 0.15 to 0.24 Hz, which is $\fn/\fpeak \sim 40$ times slower than the branch motion. If the flow stream was to buffet the coral, it might induce the back and forth gentle sway of the entire coral, but not the rapid dynamics of the branches.

\begin{figure}
\centerline{
\includegraphics[height=0.325\textheight]{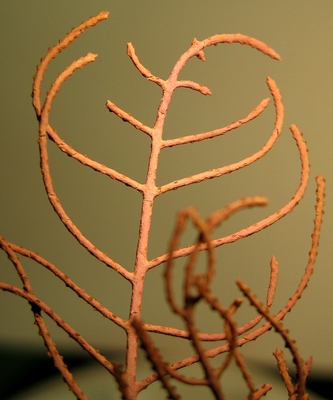}
}
\caption{Dried bipinnate sea plume \textit{Antillogorgia~bipinnata} (\citet{nova_south_eastern_university_south_2016}, \textcopyright\ Charles G. Messing). The centre-to-centre separation space between branches goes from 4.5 to 8.5 diameters. The cortex of some polyps can be discerned in the extremities of the branches (top left, bottom left and bottom right of the image).
}
\label{fig:branch}
\end{figure}

Additionally, even though the coral branches lie side-by-side in the same plane facing the flow, we do not suspect a fluidelastic instability to take place. In fact, if a pair of cylinders are separated by a centre-to-centre distance of less than $\sim 4$ diameters, they generate either a combined, a bistable asymmetric or a coupled vortex shedding, which triggers vibrations \citep{blevins_flow-induced_1990, huera-huarte_flow-induced_2011}.
We based our estimation on the dried \textit{A.~bipinnata} pictured in figure \ref{fig:branch}, and found that the separation distance was between 4.5 and 8.5 diameters. Therefore, the branches are fairly isolated from each other to shed independent vortex streets and prevent any fluidelastic instability.

A final plausible cause of vibration that might be involved is galloping. A branch cross-section is overall circular. It is known that circular cylinders are `immune to galloping' and undergo only VIVs \citep{nakamura_galloping_1994, paidoussis_fluid-structure_2010}. However, one might think that the coral branch is not perfectly circular owing to the polyps covering it. By assuming they represent small geometrical perturbations, we simulated flows around an idealised cross-section of a soft coral branch of \textit{A.~bipinnata} using an in-house flow solver \citep{etienne_perspective_2009}, as detailed in appendix~\ref{app:galloping}. We found that the Glauert-den Hartog criterion was unfulfilled, which implied that galloping is implausible as the source of soft coral branch vibrations.

In summary, turbulent buffeting, fluidelastic instability and galloping cannot be the source of the high frequency vibration of \textit{A.~bipinnata} branches, which bear all the telltale signs of VIVs.

%% file: 4_advection/advection.tex
\section{Particle advection}
\label{sec:advection}

\subsection{Coral branch as a circular spring-mounted cylinder in flow}
\label{subsec:flow}

Following the identification of VIVs as the probable source of soft coral branch vibration, we modelled the capture of particles advected by the fluid flow. We first idealised the coral branch as a circular cylinder. From a geometrical perspective, the polyps covering a branch of \textit{A. bipinnata} are small perturbations, of approximately 10\% of the branch diameter \citep{bayer_shallow-water_1961}. From a fluid-structure interaction perspective, isolated circular cylinders are only prone to VIVs \citep{nakamura_galloping_1994}, which are the flow-induced vibrations in which we are interested. Additionally, simplifying the coral branch to a circular cylinder brings our study back into the classical problem of particle capture by a circular collector in the field of particle filtering.

\begin{figure}
\centerline{
\includegraphics[width=\textwidth]{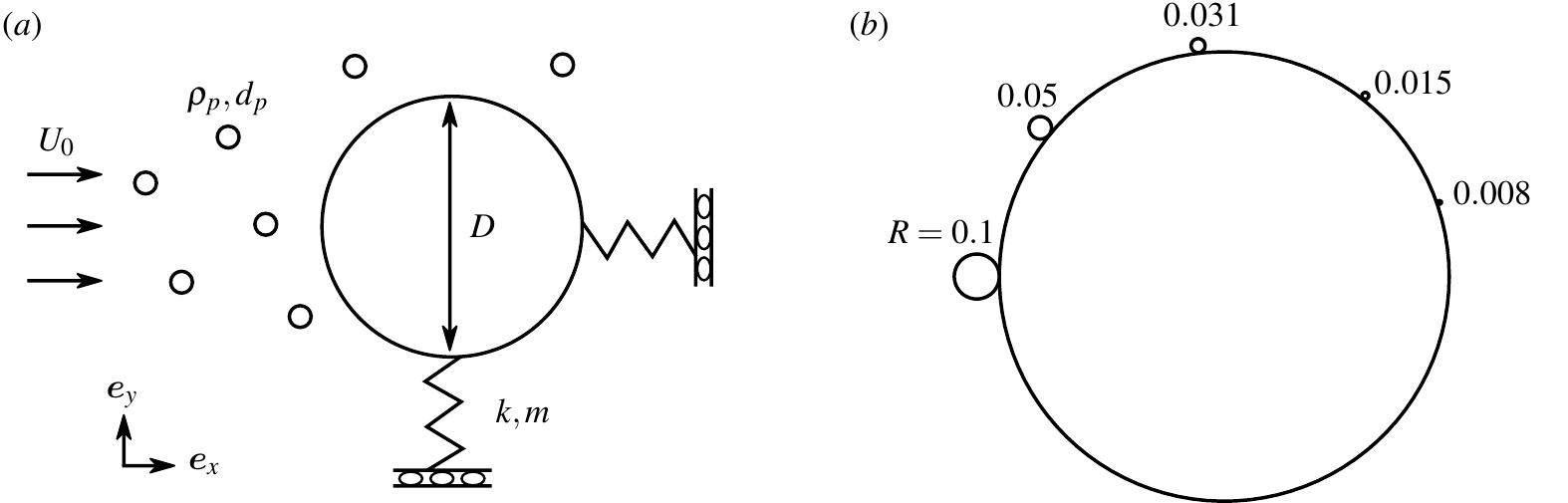}
}
\caption{($a$) Schematics of particles advected by a flow around a cylinder free to oscillate in the transverse and stream-wise directions. The upstream fluid velocity is parallel to $\ex$. Particles have a density $\rhop$ and diameter $\deep$. The mass per unit length of the cylinder is $m$, and the spring stiffness per unit length $k$ is the same in both directions. ($b$)~Drawing contrasting the size of particles simulated with the size of the cylinder, where $R = \deep/D$ is the diameter ratio.}
\label{fig:schematic-branch}
\end{figure}

We simulated a fluid flow of upstream velocity $U_{0}$ around a fixed or vibrating cylinder of diameter $D$ to highlight the effect of vibrations on particle capture. We considered the water flow as being incompressible and two-dimensional. In addition, we assumed it was unidirectional because the Keulegan-Carpenter number evaluated in the stream-wise direction of a coral branch was sufficiently large $KC \sim 250$, as estimated in section \ref{sec:origin}, and we are interested in the vortex shedding regime. The flow is governed by the continuity and Navier-Stokes equations, written in an Arbitrary Lagrangian Eulerian (ALE) framework as
\begin{subeqnarray}
&&\bnabla \cdot \bUf = 0,\\
&&\partialDeriv{\bUf}{t} + \left[ (\bUf - \bV) \cdot \bnabla \right] \bUf
= \frac{1}{\rhof} \left[ -\bnabla p + \muf \nabla^{2}\bUf \right],
\end{subeqnarray}
where $\rhof$, $\muf$, $\bUf$ and $p$ are, respectively, the fluid density, dynamic viscosity, velocity and pressure. The vector $\bV$ is the velocity of the moving mesh. If the cylinder is fixed, we simply have $\bV = \boldsymbol{0}$. The hydrodynamic load applied on the cylinder is expressed as
\begin{equation}
\bF\subrm{hydro} = \oint \left[ -p\mathsfbi{I} + \muf \left( \bnabla\bUf + \bnabla\bUf^{\textsf{T}} \right) \right]\rmdee\bl,
\end{equation}
with $\rmdee\bl$ being the integration element around the cylinder, pointing outwards. Furthermore, we considered the cylinder as a spring-mounted oscillator free to move stream-wise and transverse to the main flow under the hydrodynamic load $\bF\subrm{hydro}$, as schematised in figure~\ref{fig:schematic-branch}($a$). Neglecting structural damping, the position of the cylinder centre $\bXcyl$ is governed by
\begin{equation}
m\totalDeriv{^{2}\bXcyl}{t^{2}} + k\bXcyl = \bF\subrm{hydro},
\end{equation}
where $m = \upi\rhos D^{2}/4$ is the cylinder mass and $k$ the spring stiffness, both per unit length.

Normalising velocities with $U_{0}$ and distances with $D$, and scaling time by $D/U_{0}$ and pressure by $\rhof U_{0}^{2}$, the governing equations of the fluid and cylinder interaction can be rewritten as
\begin{subeqnarray}
&&\bnablaLess \cdot \bUfLess = 0,\\
&&\partialDeriv{\bUfLess}{\tLess} + \left[ (\bUfLess - \bVLess ) \cdot \bnablaLess \right] \bUfLess
= \left[ -\bnablaLess \pLess + \frac{1}{\Rey} \nablaLess^{2}\bUfLess \right],\\
&&\bFLess\subrm{hydro}
= \oint \left[ -\pLess\mathsfbi{I} + \frac{1}{\Rey} \left( \bnablaLess\bUfLess + \bnablaLess\bUfLess^{\textsf{T}} \right) \right]\mathrm{d}\blLess, \label{eq:force-hydro} \\
&&\totalDeriv{^{2}\bXcylLess}{\tLess^{2}}
+ \left( \frac{2\upi}{\Ur} \right)^{2} \bXcylLess = \frac{4/\upi}{M} \bFLess\subrm{hydro},
\label{eq:complete_fsi_system}
\end{subeqnarray}
with the bar $\less{(.)}$ denoting dimensionless variables. The dynamical parameters present in the system of equations (\ref{eq:complete_fsi_system}) are the Reynolds number $\Rey$, the reduced velocity $\Ur$ and the mass number $M$, which are defined as
\begin{equation}
\Rey = \frac{\rhof U_{0} D}{\muf},\quad
\Ur = \frac{2\upi U_{0}}{D} \sqrt{\frac{m}{k}},\quad
M = \frac{m}{\upi\rhof D^{2}/4}.
\label{eq:Re_Ur_M}
\end{equation}
As mentioned in section \ref{sec:origin}, we supposed that soft coral branches have a density roughly the same as water $\rhos \sim \rhof$, thus we considered a constant cylinder mass number $M = 1$ throughout this study.

\begin{figure}
\centerline{
\includegraphics[width=\textwidth]{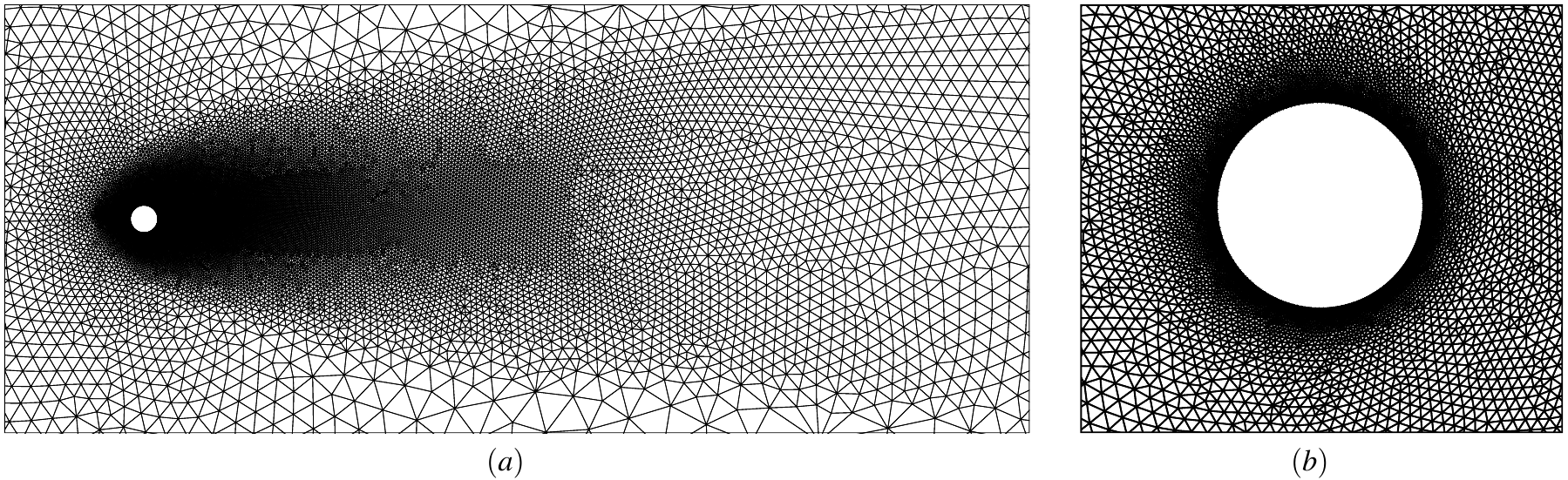}
}
\caption{Close-ups of the fluid mesh highlighting the densification in the wake (left) and around the boundary layer (right).}
\label{fig:mesh}
\end{figure}

Figure \ref{fig:mesh} shows the computational domain. The dimensions were sufficiently large ($-40 \leq \xLess \leq 120$ and $-60 \leq \yLess \leq 60$) to avoid confinement effects \citep{persillon_physical_1998}. The domain was discretised with 96 000 nodes and 48 000 Taylor-Hood ($\mathcal{P}_{2}-\mathcal{P}_{1}$) triangular elements \citep{taylor_numerical_1973}, so it follows that the velocity was third order accurate and the pressure was second order accurate. These elements were small in the wake and close to the cylinder to resolve the vortex shedding and the boundary layer.
The grid in figure~\ref{fig:mesh} was validated in \citet{etienne_low_2012} and yielded accurate results in good agreement with the existing data.
The boundary conditions on the fluid velocity $\bUfLess$ were a uniform Dirichlet at the entry ($\xLess = -40$), homogeneous Neumann at the exit ($\xLess = 120$), symmetry at the top and bottom boundaries ($\yLess = \pm 60$), and no-slip at the cylinder wall.

\subsection{Particle dynamics}
Next, we considered food particles as spheres of diameter~$\deep$ and density~$\rhop$. Several governing equations were established to describe the motion of particles in non-uniform or random flow fields \citep{maxey1983, ounis1990, barton1995}. They included terms displaying the influence of particles on the flow. However, in the present study, we assumed that the particles did not disturb the flow and could not actively change their trajectory, unlike a swimming plankton for instance. Then the suitable governing equation to consider was the Basset-Boussinesq-Oseen (BBO) equation \citep{clift_bubbles_1978, barton1995}. We neglected gravity effects, and the remaining forces applied on a particle that we kept were the drag $\Fd$, the pressure load $\Fp$ and the added mass force $\Fa$. The governing equations describing the particle trajectory are
\begin{subeqnarray}
&\displaystyle\totalDeriv{\bxp}{t} &= \bup, \\
&\displaystyle\totalDeriv{\bup}{t} &= \frac{1}{\emp} (\Fd + \Fp + \Fa),
\label{eq:particle-eq}
\end{subeqnarray}
where $\emp = \rhop\upi\deep^{3}/6$ is the mass of the particle, $\bxp$ is its position and $\bup$ is its velocity.

The drag applied on a sphere is
\begin{equation}
\Fd = -\frac{1}{2} \Cd^{\mathrm{sph}} \rhof \upi \left( \frac{\deep}{2} \right)^{2}
\norm{\bup - \bUf}
\left( \bup - \bUf \right),
\label{eq:drag}
\end{equation}
which is the same expression if the flow were steady. Here $\Cd^{\mathrm{sph}}$ is the drag coefficient of a sphere and it is a function of the particle-based Reynolds number
\begin{equation}
\Rep = \frac{\rhof \norm{\bup - \bUf} \deep}{\muf}.
\end{equation} 
Because this latter is unknown \textit{a priori}, we considered the Schiller-Nauman interpolation \citep{clift_bubbles_1978}
\begin{equation}
\Cd^{\mathrm{sph}} = \frac{24}{\Rep} (1 + \Rep^{0.687}),
\end{equation}
which is valid for $\Rep < 800$. The pressure load is related to the pressure gradient through
\begin{equation}
\Fp = - \frac{4}{3} \upi \left( \frac{\deep}{2} \right)^{3} \bnabla p.
\label{eq:pressure}
\end{equation}
It is also known as the Froude-Krylov force \citep{fredsoe_hydrodynamics_2006} owing to the difference in pressure of the global flow, which depicts the unsteadiness of the flow in the absence of the particle. If the fluid were stagnant, there would be no pressure gradient accelerating it and this force would simply vanish.

Finally, although previous studies have ignored the added mass force \citep{haugen_particle_2010, krick_adding_2015}, we deem it important to be included in the force balance. It is proportional to the particle acceleration relative to the fluid
\begin{equation}
\Fa = - \Cm \rho\subrm{f} \frac{4}{3} \upi \left( \frac{\deep}{2} \right)^{3}
\left( \totalDeriv{\bup}{t} - \totalDeriv{\bUf}{t} \right).
\end{equation}
Here $\Cm$ is the mass coefficient of a sphere. We took it to be a constant equal to 1/2. Generally, it increases rapidly when the particle comes close to the wall \citep{brennen_review_1982}, so our leading order simplification becomes invalid with proximity to the cylinder. An elaborate model would account not only for a varying mass coefficient, but also for the repulsive effect just prior to capture owing to confinement, for example by including an additional force in the momentum equation~(\ref{eq:particle-eq}) \citep{beguin_void_2016}. In all of our entire simulations, however, we kept the assumption $\Cm = 1/2$.

Denoting the ratio of diameters as
\begin{equation}
R = \frac{\deep}{D},
\end{equation} 
the particle-based Reynolds number is $\Rep = \Rey R \norm{\bupLess - \bUfLess}$, and equations (\ref{eq:particle-eq}) become
\begin{subeqnarray}
&\displaystyle\totalDeriv{\bxpLess}{\tLess} &= \bupLess, \\
&\displaystyle\totalDeriv{\bupLess}{\tLess} &= \FdLess + \FpLess + \FaLess,
\label{eq:particle-eq-dless}
\end{subeqnarray}
with
\begin{subeqnarray}
&\FdLess &= - \frac{18}{\rhoLess R^{2}\Rey}
\left( 1 + \Rep^{0.687} \right)
\left( \bupLess - \bUfLess \right), \label{eq:drag-dless} \\
&\FpLess &= - \frac{1}{\rhoLess} \bnablaLess \pLess, \\
&\FaLess &= - \frac{\Cm}{\rhoLess} \left( \totalDeriv{\bupLess}{\tLess} - \totalDeriv{\bUfLess}{\tLess} \right),
\end{subeqnarray}
where $\rhoLess = \rhop/\rhof$ is the ratio of densities. In this paper, $\rhoLess = 2$, which we considered as the upper bound of densities of the existing food particles. According to \citet{espinosa-gayosso_density-ratio_2015}, a sediment-type particle that is 2.6 times heavier than water achieves a capture efficiency equal to that of a perfect fluid tracer if the Stokes number $Stk = \rhoLess R^{2} \Rey/9$ is less than 0.1 at $\Rey \sim 100$. For a typical simulation case at this Reynolds number we have $Stk \sim 0.06$ (e.g. taking $R \sim 0.05$), therefore, the choice $\rhoLess = 2$ is legitimate.

\subsection{Numerical solving}
\label{subsec:solving}
We solved the particle advection problem in two stages. First, we computed the FSI problem in an ALE framework using the in-house solver
\textsc{Cadyf} \citep{etienne_perspective_2009}. This code was verified with the method of manufactured solutions \citep{hay2014, yu2015} and for classical FSI problems such as airfoil plunging and pitching and flexible strip at the rear of a square cylinder \citep{cori2015}. The code was also validated and produced dynamical results consistent with the data in the literature for several flow-induced vibrations including VIVs of circular \citep{etienne_low_2012} and square \citep{hay2015} cylinders, and wake-induced vibrations of cylinders in tandem \citep{yu2016}.

Initially, the fluid was at rest, $\bUfLess = \boldsymbol{0}$ and $\pLess = 0$, and ramped up to unity $\bUfLess = \ex$ in a few time steps to satisfy a divergence-free flow field. Also at the initial time, the cylinder stood at the position $\bXcylLess = \boldsymbol{0}$. \textsc{Cadyf} integrated the coupled system of equations (\ref{eq:complete_fsi_system}) using hp-adaptive backward differential formulas (BDF) methods \citep{hay_hp-adaptive_2015}. The order and time step adjust automatically so that the local truncation error remained smaller than a constant absolute tolerance equal to $10^{-5}$. We extracted the flow solution at each instant $\tLess_{n} = n/10$ starting from $\tLess = 400$, long after VIVs had reached their periodic limit-cycle.

Next, we exported these data into \citet{mou3adb_paradvect_2020}, a Python code we wrote in order to integrate the system of equations (\ref{eq:particle-eq-dless}) in a Lagrangian framework using a forward Euler scheme. A code verification of \textsc{Paradvect} is available in the other supplementary material. To satisfy stability, the integration time step~$\Delta \tLess$ should be smaller than the characteristic time involved in the system $\tau = (\rhoLess + \Cm) R^{2}\Rey/9$.
We compared $\tau$ with the time step $\tLess_{n+1} - \tLess_{n} = 1/10$ by which we extracted the flow solution from \textsc{Cadyf}: if $\tau > 1/10$, which is the case for large particles, we took $\Delta \tLess = 1/10$, whereas if $\tau < 1/10$, which is the case for small particles, we linearly interpolate the flow solution in time.
Linear interpolation is appropriate because the flow solution varies smoothly during a time step of $\sim 1/10$ and does not fluctuate over the small timescale $\tau$.

Figure~\ref{fig:schematic-branch}($b$) juxtaposes the cylinder and the considered particles to give an idea on their relative size. The trajectory of every particle was integrated from a unique starting line~$\xLess_{0} = 2$ upstream from the cylinder for several ordinates $\yLess_{0}$. The flow in these positions was horizontal and unperturbed. The local fluid velocity was then assigned as their initial velocity.

To calculate the total hydrodynamic force applied on the particle, we quadratically interpolated the fluid velocity and linearly interpolated the pressure at the particle centre. As this calculation requires knowledge of the finite element hosting the particle, we adopted the particle tracer algorithm proposed by \citet{lohner_vectorized_1990}. It searches recursively, neighbour to neighbour, the new host element in the vicinity of the known previous host element. This tracking technique is suitable, time-saving and easy to implement in particle-in-cell codes, such as ours, where physical particles do not jump over many elements in a single time step \citep{lohner_applied_2008}.

Finally, we assumed that the particle was captured as soon as it hit the edge of the cylinder. Because the particles were spherical and the cylinder was circular, the capture occurred when $\norm{\bxpLess-\bXcylLess}~\le(1+R)/2$. We refer to this capture condition as the \textit{solid contact} criterion.

%% file: 5_capture/capture.tex
\section{Capture rate}

\subsection{Definition}
\begin{figure}
\centerline{
\includegraphics[width=\textwidth]{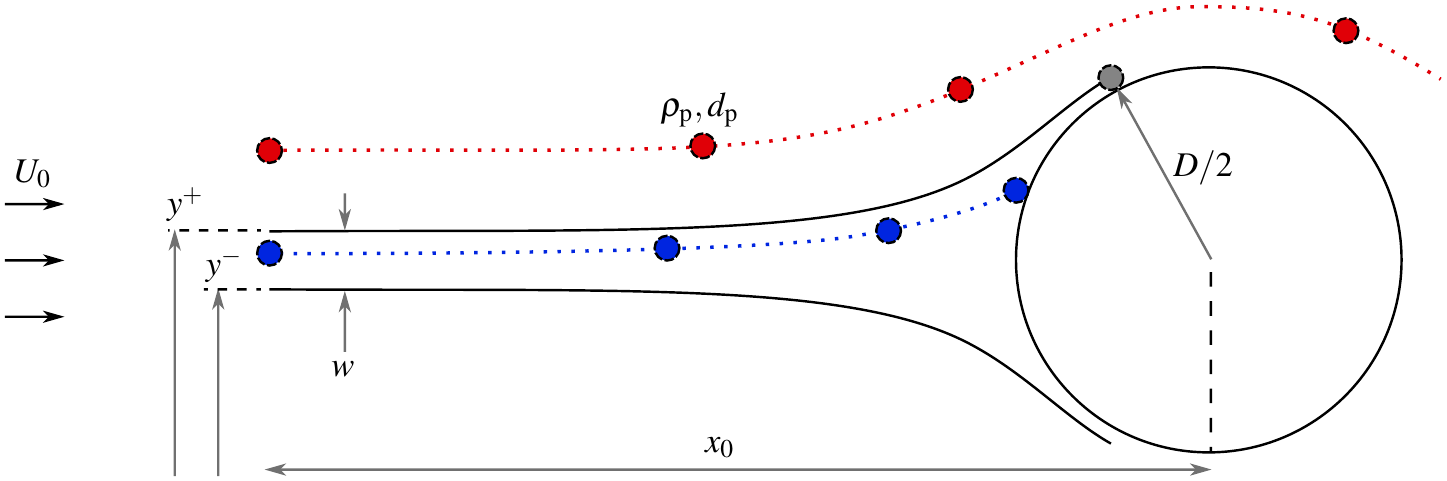}
}
\caption{Schematics of particle interception by a fixed cylinder. Particles are launched from a distance $x_{0}$ upstream from the cylinder. Because the blue particle starts near the symmetry line, it is captured, whereas the red particle starts from a higher $y$-position and succeeds in crossing over the cylinder and escaping capture. The grey particle is the farthest particle from the symmetry line that the cylinder intercepts. Its trajectory starts from the ordinate $y^{+}$, and defines the upper border of the capture domain. The size of the capture window is $w = y^{+} - y^{-}$.}
\label{fig:capture_domain_fixed}
\end{figure}

To assess the filtering ability of the cylinder, we calculated the rate at which it intercepts particles. We defined the capture rate $\Ndot$ as the number of particles that the cylinder captures per unit time. As represented in figure \ref{fig:capture_domain_fixed}, the particle that would be ultimately captured necessarily entered through an opening that we called the \textit{capture window}. Thereby, the capture rate also equals the flux of particles through this capture window
\begin{equation}
\Ndot =  C_{0}U_{0}w,
\label{eq:def_Ndot}
\end{equation}
where $C_{0}$ is the particle concentration per unit length, which we assume constant and uniform, and $w$ is the size of the capture window.

\subsection{Calculation strategy}
\subsubsection{Automated dichotomy}
Owing to the definition (\ref{eq:def_Ndot}), we determined the capture rate through the calculation of the size of the capture window. The upper and lower boundaries of the capture region were the trajectories of the farthest captured particles, as shown in figure~\ref{fig:capture_domain_fixed}. Thus, it suffices to calculate the initial ordinates of these border particles, denoted $y^{+}$ and $y^{-}$, because
\begin{equation}
w = y^{+} - y^{-}.
\end{equation}
First, we considered a captured particle (in blue, figure~\ref{fig:capture_domain_fixed}) released from ($x_{0}, \yc$), and a non-captured particle (in red) released from ($x_{0}, \ync$). Their initial positions were necessarily bounds of either border, say the upper border~($\yc < y^{+} < \ync$). Next, we released a third particle between the two previous ones, from the middle of their initial ordinates $\ym = (\yc + \ync)/2$. If it escaped capture, then the upper border $y^{+}$ was necessarily between the initial ordinates of this new particle and the former captured one ($\yc < y^{+} < \ym$). The code \citet{mou3adb_paradvect_2020} repeated this dichotomic process until it reached a resolution of one thousandth of the particle diameter $| \yc - \ym | < \deep/1000$.

\subsubsection{Temporal decomposition}
\begin{figure}
\centerline{
\includegraphics[width=\textwidth]{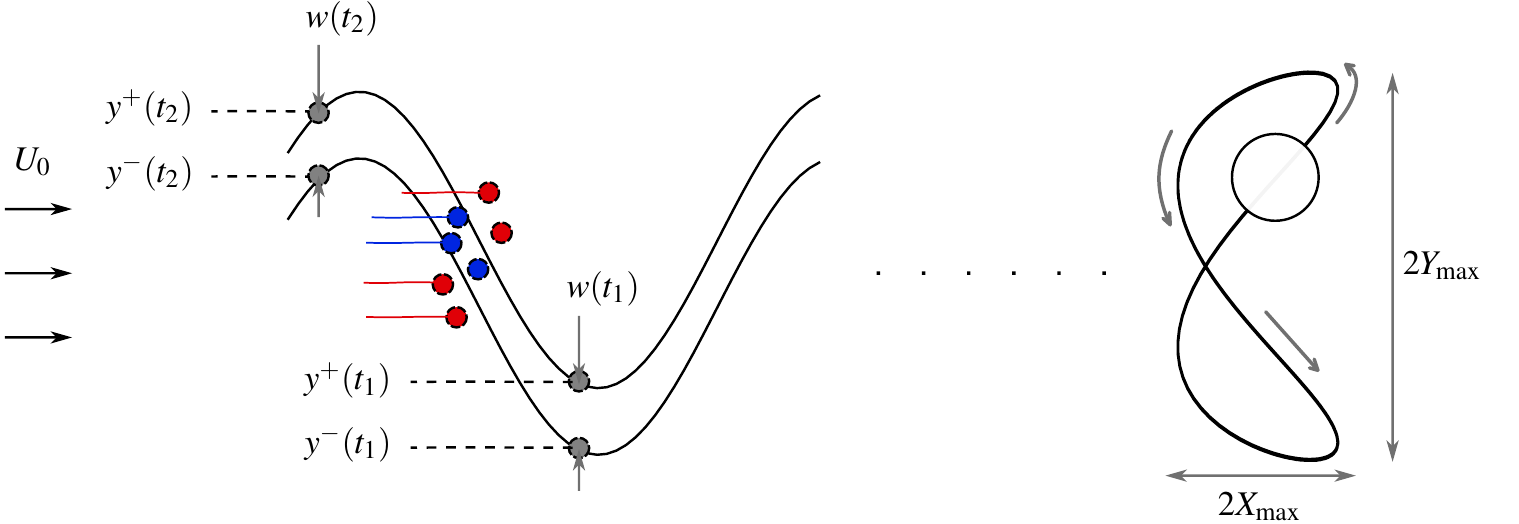}
}
\caption{
Distortion of the capture domain owing to vibration. The initial ordinates of the grey border particles are now varying in time $y^{\pm} = y^{\pm}(t)$. The instantaneous capture window, through which the particles launched at time $t$ would be captured, is $w(t) = y^{+}(t) - y^{-}(t)$. At instants $t$ between $t_{1}$ and $t_{2} > t_{1}$, the blue particles are launched between $y^{-}(t)$ and $y^{+}(t)$, so they are captured. The red particles are launched outside of these limits, so they escape capture. A portion of particle trajectories is shown by the solid line.
The cylinder draws a lemniscate trajectory, and $\Ymax$ and $\Xmax$ are the transverse and stream-wise amplitudes of vibration. The motion of the cylinder is symmetric with respect to the horizontal line passing through the node of the lemniscate, hence the capture is periodic and goes in harmony with the stream-wise oscillations, which complete a cycle in half a period.
}
\label{fig:capture_domain_vibrating}
\end{figure}

When the cylinder vibrates, the borders of the capture region vary in time $y^{\pm} = y^{\pm}(t)$, and the capture strip is no longer straight, as seen in figure~\ref{fig:capture_domain_vibrating}. A particle would be captured if it is launched at time $t$ between $y^{+}(t)$ and $y^{-}(t)$. The instantaneous capture window $w(t) = y^{+}(t) - y^{-}(t)$ must be periodic, thus we propose the following ansatz
\begin{equation}
w(t) = \wmean + \wamp \sin\left( \frac{2\pi t}{T} + \varphi \right).
\label{eq:ansatz_w}
\end{equation}
The period $T$ is not equal to the vibration period, but half of it. Indeed, as illustrated in figure~\ref{fig:capture_domain_vibrating}, the cylinder describes a lemniscate, so it captures particles in the same way whether during the upper or the lower loop. Therefore, the capture rate is determined by the stream-wise -- not transverse -- oscillations, which complete a cycle in half a period.

We should mention that the dependency on time is also true for the fixed cylinder, albeit with a much smaller fluctuation. For the fixed cylinder, the variation of $w(t)$ arises from the periodic vortex shedding, which affects the streamlines in the front, and the period $T$ in the ansatz~(\ref{eq:ansatz_w}) is equal to the vortex shedding period.

We calculated the size of the capture window at instants $t =$ 0, $T/4$ and $T/2$, and determined the mean and fluctuating components $\wmean$ and $\wamp$ in equation~(\ref{eq:ansatz_w}) from the ensuing formulae
\begin{subeqnarray}
&\wmean &= \frac{w(0) + w(T/2)}{2},\\
&\wamp &= \sqrt{
\left[ w(0) - \wmean \right]^{2}
+ \left[ w(T/4) - \wmean \right]^{2}
}. \label{eq:components_Ndot}
\end{subeqnarray}

\subsection{Dimensional analysis}
The capture rate is a function of the fluid flow, cylinder and particle properties
\begin{eqnarray}
&\Ndot\ [\mathrm{s^{-1}}],\quad
U_{0}\ [\mathrm{m\ s^{-1}}],\quad
\rhof\ [\mathrm{kg\ m^{-3}}],\quad
\muf\ [\mathrm{kg\ m^{-1}s^{-1}}],\quad
D\ [\mathrm{m}],\quad
k\ [\mathrm{N\ m^{-2}}],\quad& \nonumber \\
&m\ [\mathrm{kg\ m^{-1}}],\quad
\deep\ [\mathrm{m}],\quad
\rhop\ [\mathrm{kg\ m^{-3}}],\quad
C_{0}\ [\mathrm{m^{-2}}].&
\end{eqnarray} 
For convenience, we discarded the variation in time $t$ owing to the decomposition in equation~(\ref{eq:ansatz_w}). The Buckingham $\Pi$ theorem \citep{buckingham_physically_1914} states that a relation exists between $10 - 3 = 7$ independent dimensionless variables, which we choose to be
\begin{eqnarray}
&\eta = \displaystyle\frac{\Ndot}{C_{0}U_{0}D},\quad
\Rey = \frac{\rhof U_{0} D}{\muf},\quad
R = \frac{\deep}{D},\quad
\Ur = \frac{2\upi U_{0}}{D} \sqrt{\frac{m}{k}},& \nonumber \\
&\rhoLess = \displaystyle\frac{\rhop}{\rhof},\quad
M = \frac{m}{\rhof D^{2}},\quad
\less{C}_{0} = C_{0}D^{2}.&
\label{eq:dimensionless_numbers}
\end{eqnarray} 
In this way, the dimensionless capture rate $\eta$ is also the capture efficiency defined by \citet{weber_interceptional_1983}, \citet{palmer_observations_2004}, and \citet{espinosa-gayosso_particle_2012, espinosa-gayosso_particle_2013}, which itself, from the definition (\ref{eq:def_Ndot}), equals the dimensionless size of the capture window
\begin{equation}
\eta = \frac{\Ndot}{C_{0}U_{0}D}  = \frac{w}{D} = \wLess.
\label{eq:eta_is_eLess}
\end{equation}
According to equation~(\ref{eq:ansatz_w}), it follows that 
\begin{equation}
\eta = \etaMean + \etaAmp \sin\left( \frac{2\pi t}{T} + \varphi \right),
\label{eq:ansatz_eta}
\end{equation}
with $\etaMean = \wmean/D$ and $\etaAmp = \wamp/D$.
For all simulations in the present work, we kept a constant and uniform particle concentration $\less{C}_{0} =$ const. In addition, as we assumed in section~\ref{sec:advection}, we chose $M = 1$ and $\rhoLess = 2$.
Henceforth, the remaining variables that determine the capture rate are the Reynolds number $\Rey$, the diameter ratio $R$ and the reduced velocity $\Ur$.

\section{Results}
To see how the vibration affects the capture rate, we allowed the cylinder to freely oscillate and varied the spring stiffness so that the reduced velocities were between $\Ur = 1$ and 13, which covered the estimation made in section~\ref{sec:origin}. We simulated the trajectories of particles having diameter ratios $0.015 \le R \le 0.1$, the same range as for phytoplankton and larvae compared with \textit{A. bipinnata} branches \citep{shimeta_physical_1991}. The considered flows had Reynolds numbers ranging from $\Rey = 50$ to 200, which were the respective limits of the establishment of the laminar von K\'{a}rm\'{a}n vortex street and the transition into three-dimensional vortex shedding.

In equation~(\ref{eq:ansatz_eta}), we found a ratio $\etaAmp/\etaMeanF$ of less than 2\% for the fixed case, which meant that the temporal term could be regarded as a small fluctuation. For the vibrating case, this ratio was larger $\etaAmp/\etaMean \lesssim 20\%$. We did not investigate the time fluctuations of the particle capture rate, but instead focused on the time-averaged capture rate because it is most relevant in terms of filtering applications and biological implications.

\begin{figure}
\centerline{
\includegraphics{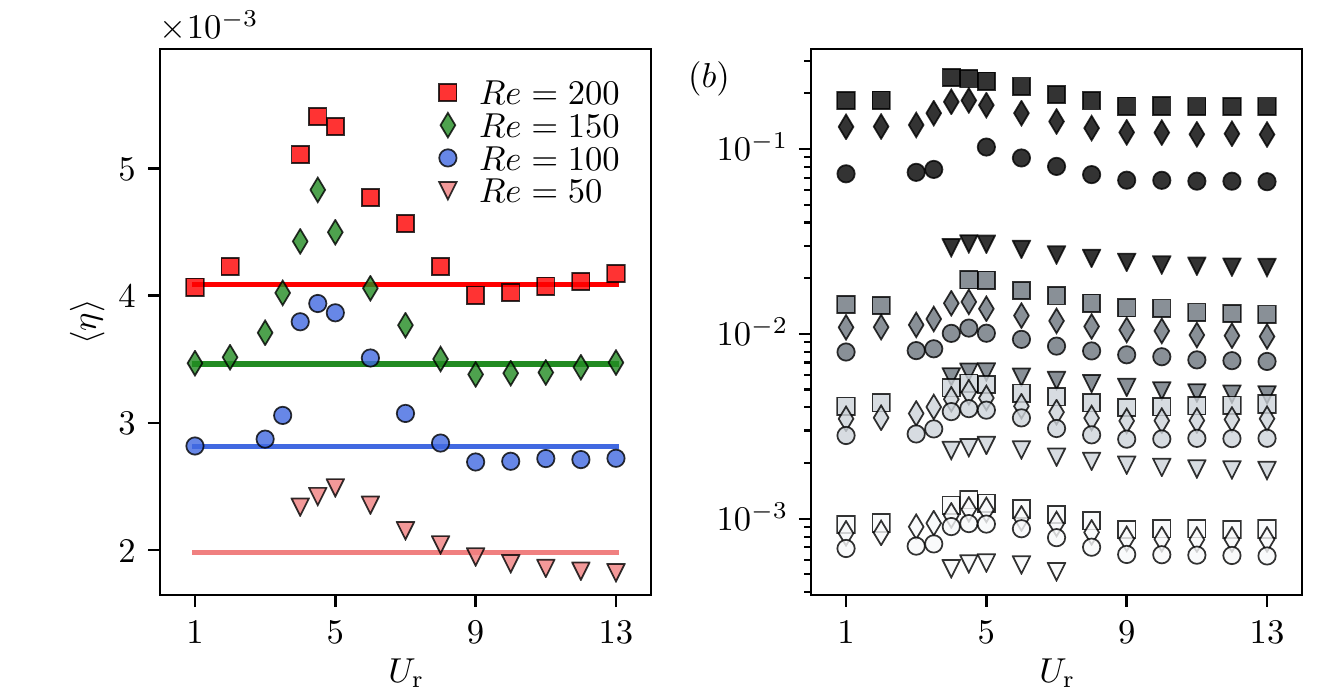}
}
\caption{Mean capture rate $\etaMean$ versus the reduced velocity for ($a$) the particle of diameter ratio $R = 0.031$, and for ($b$) all particles $0.015 \le R \le 0.1$ in flows of Reynolds numbers $50 \le \Rey \le 200$. In ($b$), the $y$-axis is logarithmic, and each group of four curves with a single grey level corresponds to the particles $R = 0.015$ (white), 0.031 (light grey), 0.05 (dim grey) and 0.1 (black).}
\label{fig:eta_vs_Ur}
\end{figure}

\begin{figure}
\centerline{
\includegraphics{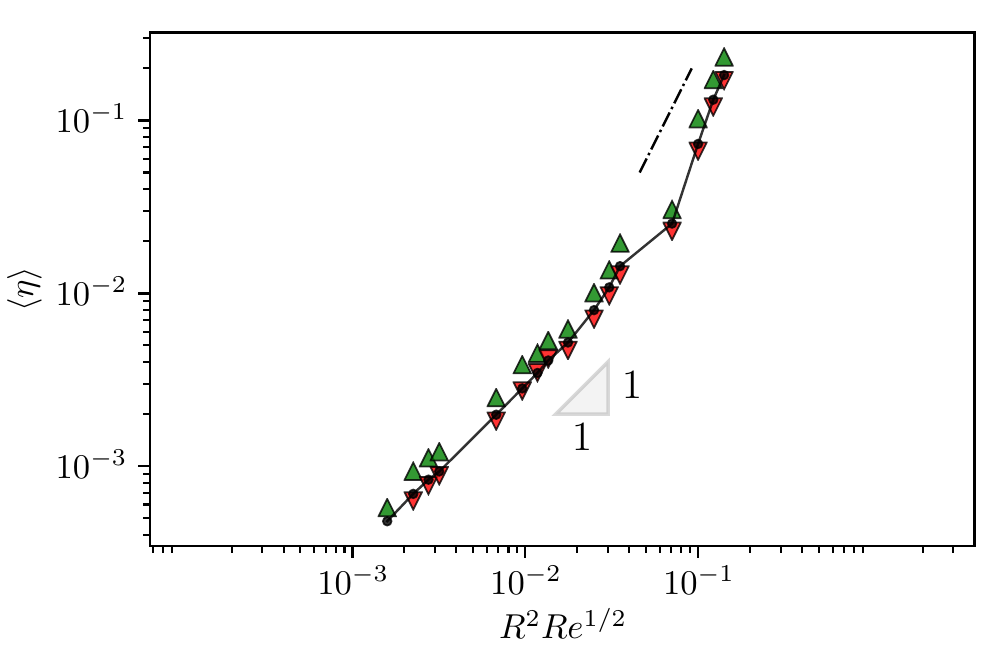}
}
\caption{
Mean capture rate of the fixed cylinder (\textcolor{black}{$\fullcirc$}) and the vibrating cylinder (\textcolor{green}{$\blacktriangle$}, $\Ur = 5$ and \textcolor{red}{$\fulltriangledown$}, $\Ur = 13$) versus the product $R^{2}\Rey^{1/2}$. For $R^{2}\Rey^{1/2} \lesssim 0.06$, the capture rate varies as $\etaMean \sim \etaMeanF \sim R^{2}\Rey^{1/2}$. Beyond this value, the scaling deviates to $R^{2}\Rey$, as indicated by the dash-dotted line (\protect\mbox{-- $\cdot$ --}).
}
\label{fig:eta_vs_product}
\end{figure}

Figure \ref{fig:eta_vs_Ur} shows the variation of the mean capture rate~$\etaMean$ with the reduced velocity~$\Ur$. We see that~$\etaMean$ was a bell-shaped function, which started from almost the same value as for a fixed cylinder ($\Ur \sim 1$), peaked at lock-in ($\Ur \sim 5$), then decreased down to values either around or less than the capture rate by a fixed cylinder.
We observe in figure~\ref{fig:eta_vs_Ur}($a$) that~$\etaMean$ increased with the Reynolds number $\Rey$, and from figure~\ref{fig:eta_vs_Ur}($b$) it increased with the particle size $R$ as well. Figure~\ref{fig:eta_vs_product} shows the variation of the mean capture rate with these two parameters for a given reduced velocity $\Ur$. We found that the mean capture rate followed the same scaling as in the fixed cylinder case $\etaMean \sim \etaMeanF \sim R^{2}\Rey^{1/2}$. The square root of the Reynolds number was reminiscent of the boundary layer thickness on the cylinder wall. The boundary layer played a key role in the capture process \citep{haugen_particle_2010}, and explained the $R^{2}\Rey^{1/2}$ power law \citep{boudina2020}.

\begin{figure}
\centerline{
\includegraphics{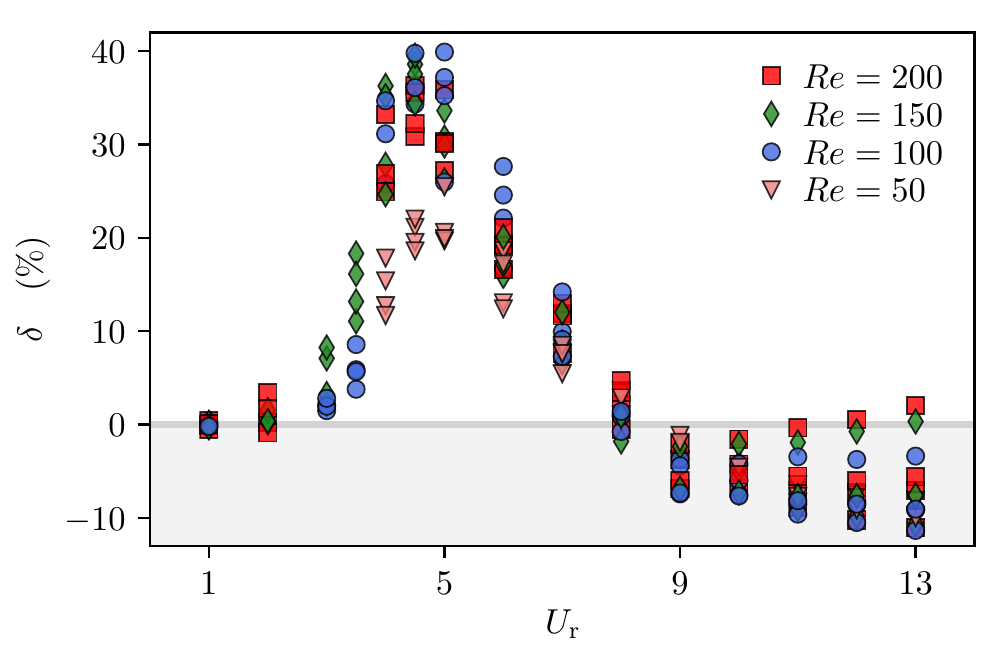}
}
\caption{Variation of the gain in capture rate versus the reduced velocity for all simulation cases $0.015 \le R \le 0.1$ and $50 \le \Rey \le 200$.}
\label{fig:delta_vs_Ur}
\end{figure}

To evaluate the benefit that vibrations bring to particle interception, we defined the gain in the capture rate $\delta$ as the relative difference between capture rates by a vibrating and a fixed cylinder
\begin{equation}
\delta = \frac{\etaMean - \etaMeanF}{\etaMeanF}.
\end{equation}
As shown in figure \ref{fig:delta_vs_Ur}, the gain was also a bell-shaped function. In table~\ref{tab:peak_values}, we give the values of the maximum gain for each Reynolds number. The peak started from 25\% for $\Rey = 50$, and reached values of 36-40\% for $\Rey =$ 100, 150 and 200.

\begin{figure}
\centerline{
\includegraphics{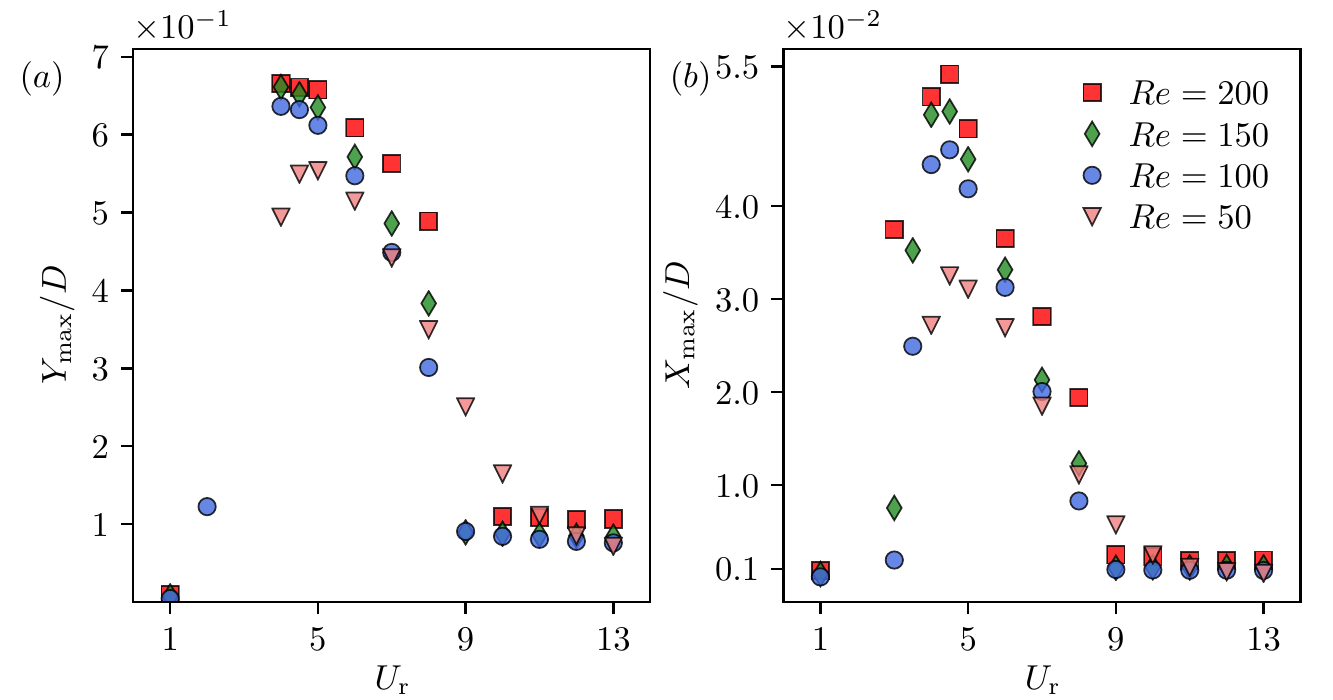}
}
\caption{Numerical response of $(a)$ the transverse and ($b$) the stream-wise amplitudes of the cylinder.}
\label{fig:Ymax_Xmax_vs_Ur}
\end{figure}

The cylinder motion had an appreciable effect on particle capture. The $\delta(\Ur)$ curve shows resemblance with the lock-in amplitude response curve \citep{paidoussis_fluid-structure_2010}. In fact, from a kinematic perspective, the curves of $\delta$ had the same profile as the responses of the transverse and stream-wise amplitudes of the cylinder $\Ymax$ and $\Xmax$, as shown in figure \ref{fig:Ymax_Xmax_vs_Ur}. These amplitudes also started from zero for small reduced velocities, peaked at lock-in, then decreased and saturated for large $\Ur$. 

\begin{table}
  \begin{center}
\def~{\hphantom{0}}
  \begin{tabular}{lcccc}
Reynolds number & 50 & 100 & 150 & 200 \\
Maximum gain (\%) & 25.5 & 39.9 & 39.5 & 36.3
  \end{tabular}
  \caption{The peak value of the gain in capture $\delta$ for each Reynolds number $\Rey$.}
  \label{tab:peak_values}
  \end{center}
\end{table}

\begin{figure}
\centerline{
\includegraphics{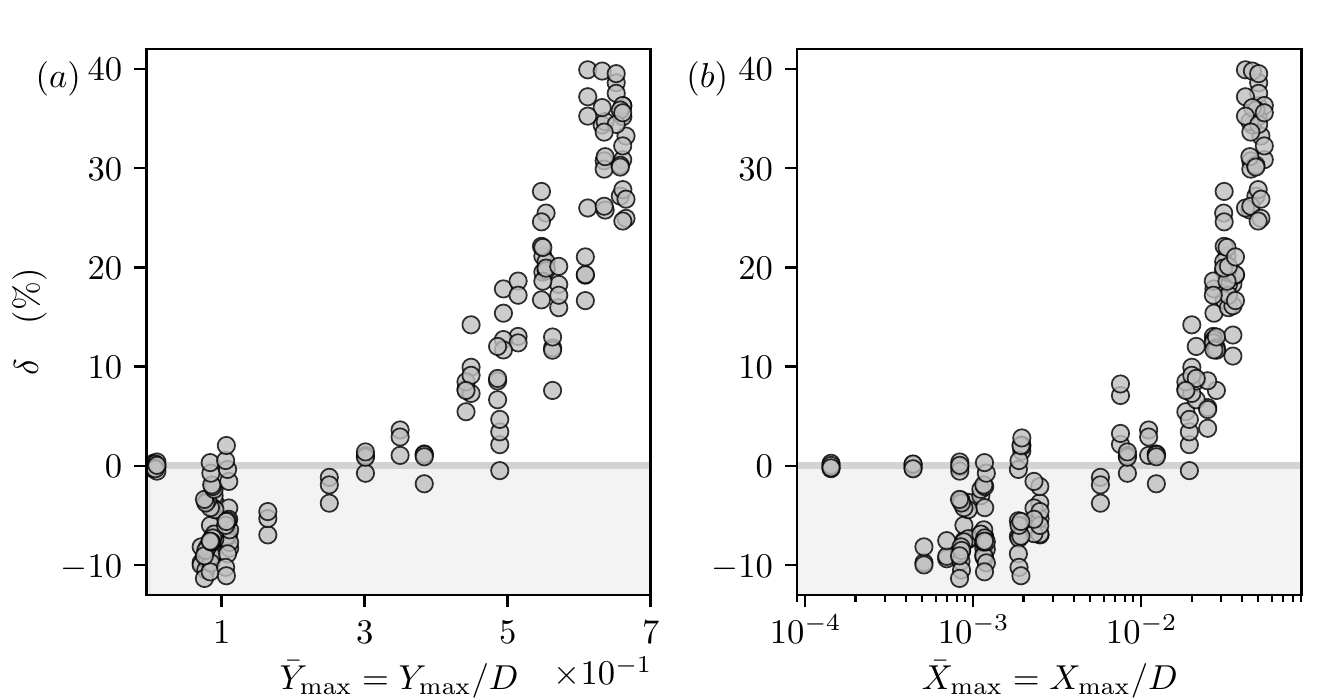}
}
\caption{Variation of the gain in capture rate versus the transverse and stream-wise amplitudes of the cylinder for all simulation cases $0.015 \le R \le 0.1$ and $50 \le \Rey \le 200$. The $x$-axis in ($b$) is logarithmic to visualise better the staggered data points.}
\label{fig:delta_vs_Ymax_Xmax}
\end{figure}

In figure \ref{fig:delta_vs_Ymax_Xmax}, the gain $\delta$ is plotted against $\YmaxLess=\Ymax/D$ and $\XmaxLess=\Xmax/D$. Beyond certain amplitude thresholds $\YmaxLess \gtrapprox 0.3$ and  $\XmaxLess\gtrapprox 0.01$,  $\delta$ is an increasing function of the amplitudes. This result is intuitive because the cylinder filters a wider cross-flow space for large $\YmaxLess$ and encounters more particles. Also, a cylinder with a large $\XmaxLess$ would have more space to accelerate against the stream and reach important counter-current velocities, which increases the speed of particles relative to the cylinder frame and subsequently the capture rate.
For $\YmaxLess > 0.2$ and $\XmaxLess > 0.01$, the results collapse into a single increasing curve (in particular, they vary linearly with $\Xmax$).

In figure \ref{fig:delta_vs_Ymax_Xmax}, it is worth noting that, past the lock-in peak ($\Ur > 9$), the vibration of the cylinder becomes detrimental to capture. This zone corresponds to the amplitudes $0.1 < \YmaxLess < 0.3$ and $10^{-3} < \XmaxLess < 0.01$.
In figure \ref{fig:delta_vs_Ymax_Xmax}, this detrimental region has data points of $\delta$ that are staggered between -10\% and 0\% around $\YmaxLess = $ $0.1$ and $\XmaxLess = 10^{-3}$, instead of a clear monotonic variation.
One way to explain this drop in capture is that while $\Ur$ increases, the cylinder not only spans short transverse distances, but it also slows down and its period becomes larger than the characteristic time of the particle advection. The particles would then see a cylinder that switches places up and down slowly, staying a long time in either side, hence giving them the opportunity to escape capture. For this reason, the cylinder misses several interception events, so it would be better if it stayed fixed, or at least in a quasi-steady state $\Ur \gtrsim 20$ \citep{blevins_flow-induced_1990}. From this viewpoint, we expect the general profile of the curves $\delta(\Ur)$ would have an ascending phase ($0 < \Ur < 5$), a peak at lock-in ($\Ur \approx 5$), a descending phase ($5 < \Ur < 9$), a detrimental regime below zero ($9 < \Ur < 20$) and a plateau towards zero beyond the quasi-steady state ($\Ur > 20$).

\begin{figure}
\centerline{
\includegraphics{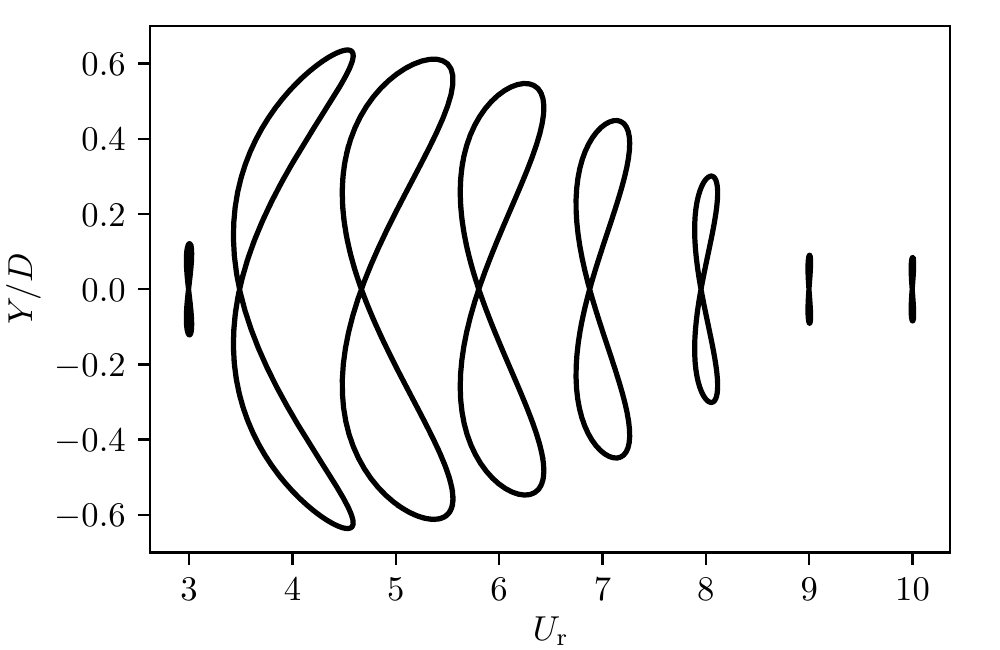}
}
\caption{Shape of the lemniscate limit-cycle trajectory of the cylinder in a flow at $Re = 100$ versus the reduced velocity. The $x$-span of each lemniscate is magnified approximately 13 times to elucidate the loops.}
\label{fig:lemniscate_vs_Ur}
\end{figure}

The matching between the gain $\delta$ and the amplitudes of the cylinder is emphasised if we consider the \textit{slenderness} ratio of its lemniscate limit-cycle trajectory defined as
\begin{equation}
\gamma = \frac{\Ymax}{\Xmax}.
\end{equation} 
Whereas $\Ymax$ and $\Xmax$ give the dimensions of the lemniscate, $\gamma$ informs on its slenderness: a small $\gamma$ illustrates an extended lemniscate and a large $\gamma$ illustrates a thin lemniscate. From figure~\ref{fig:lemniscate_vs_Ur}, we see that the lemniscate is extended around lock-in $\Ur \sim 5$ and is thin outside. 

\begin{figure}
\centerline{
\includegraphics{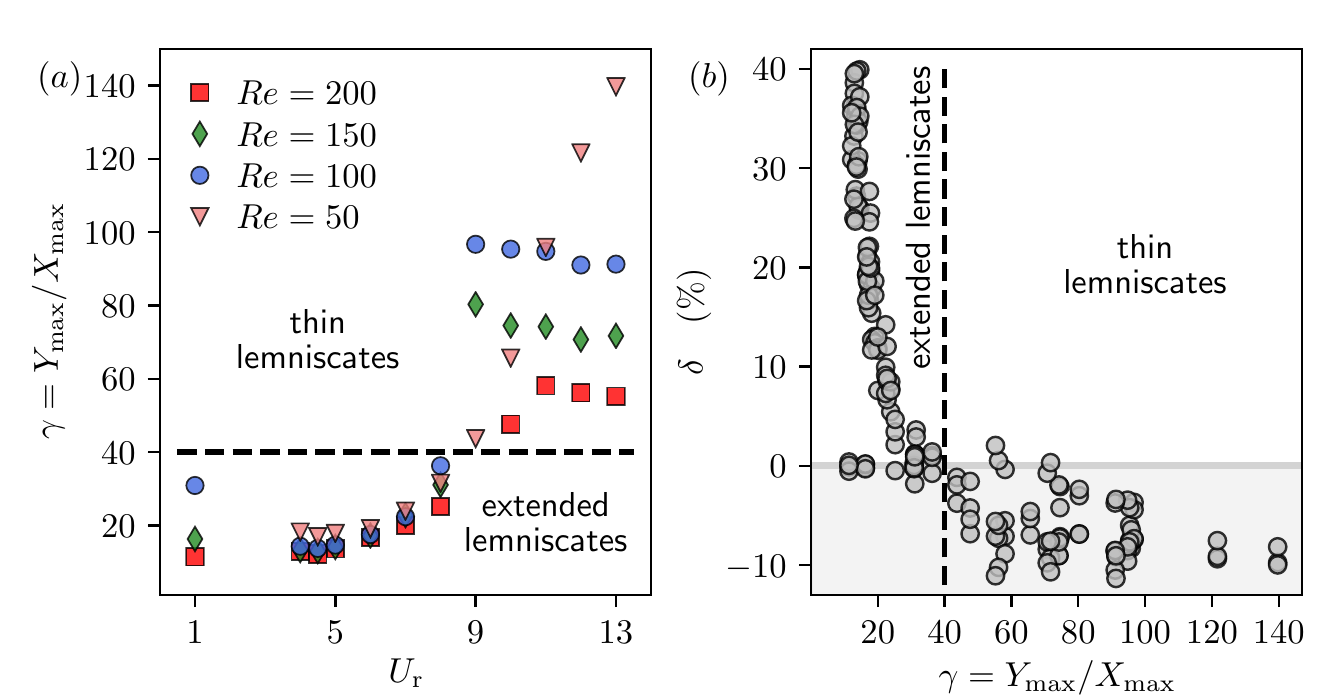}
}
\caption{($a$) Slenderness ratio of the lemniscate limit-cycle trajectory of the cylinder versus the reduced velocity for the Reynolds numbers $50 \le \Rey \le 200$. For low reduced velocities, $\gamma$ starts from a relatively small value and decreases until it reaches the minimum at lock-in ($\Ur \approx 5$). From $\Ur = 1$ to 8, the slenderness ratio for all Reynolds numbers follows the same trend. Beyond $\Ur = 8$, it keeps increasing for $\Rey = 50$, while it plateaus for $100 \le \Rey \le 200$.
($b$) Gain in capture rate versus the slenderness ratio of the lemniscate for all simulation cases $0.015 \le R \le 0.1$ and $50 \le \Rey \le 200$. For $\gamma < 40$, data points converge around two branches. The lower branch, which is close to the zero value, corresponds to low reduced velocities, whereas the upper branch represents the region around lock-in. The dashed line in both graphics is the critical value $\gamma^{*} = 40$ cutting the domain in two regions: extended lemniscates for $\gamma < 40$ and thin lemniscates for $\gamma > 40$.}
\label{fig:delta_vs_gamma_and_response}
\end{figure}

Figure \ref{fig:delta_vs_gamma_and_response}($b$) shows the graph of $\delta$ versus $\gamma$. We see that there is a critical slenderness $\gamma^{*} \approx 40$ that splits the domain into extended lemniscates ($\gamma < 40$) and thin lemniscates ($\gamma > 40$). In the region of extended lemniscates, the vibration is beneficial, and the data points collapse well around two branches: a lower branch ($\delta \gtrapprox 0$) corresponding to the small reduced velocities, and an upper branch corresponding to lock-in. In the region of thin lemniscates, the vibration is detrimental, and the data points scatter. This lack of correlation in the detrimental range is also observed in the response of the slenderness itself, as noticed in figure \ref{fig:delta_vs_gamma_and_response}($a$). Indeed, while $\gamma$ preserves the same values irrespective of the Reynolds number in the beneficial range ($1 \le \Ur \le 8$ and $\gamma < 40$), it either keeps increasing for $\Rey = 50$ or plateaus towards different values for $100 \le \Rey \le 200$ in the detrimental range ($\Ur > 9$ and $\gamma > 40$). Hence, it appears that the response of $\gamma$ dictates the outcome of vibrations: whenever the slenderness preserves a unique variation irrespective of the Reynolds number, the vibration is beneficial and the gain $\delta$ follows a clear trend, otherwise the vibration is detrimental and the variation of the gain $\delta$ remains unclear. That is to say, the shape of the lemniscate gives a clue about the impact of vibrations on capture.

%% file: 6_conclusion/conclusion.tex
\section{Biological insight}
\label{sec:biological}

From our simulations, we have seen that a rigid spring-mounted cylinder under VIVs can capture up to 40\% more particles than its fixed counterpart in an optimal range of reduced velocities, which corresponds to the lock-in range $\Ur \sim 5$.
From a biological point of view, a 40\% increase of food availability is significant, however, it occurs over a defined range of flow speeds, cylinder diameters and cylinder frequencies. The existence of an optimal capture rate implies that the flow and cylinder properties must combine in such a way that $\Ur$ remains around the peak and avoids the tails of the bell-shaped curve of $\delta$ versus $\Ur$.

If we consider the more realistic case of an elastic rod subjected to flow, we can expect two sources of additional complexity: a continuous system ($i$) vibrates with mode shapes and ($ii$) possesses many natural frequencies. Whereas VIVs of the spring-mounted cylinder involve a motion of the entire collector, only a fraction of a continuous system vibrates when it is excited in one mode shape. Then, one could think that a continuous three-dimensional system would limit the gain in capture. However, in reality, we expect that the curve of the gain in capture $\delta(\Ur)$ would comprise several peaks around multiple optimal lock-in regions $\Ur^{\mathrm{opt}}$ associated with the natural frequencies of the continuous system. Thus, once $\delta$ decreases after the first peak, it shall increase again and reach the following peak, and so forth for each $\Ur^{\mathrm{opt}}$ without finding a gap to drop below zero. We deduce that having multiple natural frequencies ensures that a soft coral colony can achieve lock-in in varying environments.

These continuous system considerations imply that soft corals could benefit from tuning their morphology (e.g. length and diameter of branches, size of polyps) as well as their structural properties (e.g. flexural rigidity of the skeleton) according to the typical ambient water speed. They could tune their properties to match one of the optimal reduced velocities $\Ur^{\mathrm{opt}}$, trigger VIVs and achieve the best feeding rate.


Let us assimilate the soft coral branch to an elastic rod having a diameter $D$, length $L$, mass per unit length $\ml$ (including the added mass) and flexural rigidity $EI$. The reduced velocity would be
\begin{equation}
\Ur = St \frac{U_{0}}{D} \sqrt{\frac{\ml L^{4}}{EI}},
\end{equation}
where $St$ is the Strouhal number. Here $\Ur$ equals the ratio of the vortex shedding frequency $St U_{0}/D$ to the characteristic frequency of the structure $\sqrt{EI/\ml L^{4}}$. Considering that $I \sim D^{4}$ and writing $\ml \sim \rho D^{2}$, we see that a reduced velocity sticking around the optimal range should verify
\begin{equation}
St U_{0} \left( \frac{\rho}{E} \right)^{1/2} \frac{L^{2}}{D^{2}} \sim \Ur^{\mathrm{opt}}\ (= \mathrm{const}).
\label{eq:matching_optimal}
\end{equation}
Introducing the aspect ratio $\Gamma = L/D$ and the speed (of sound) $\cs = \sqrt{E/\rho}$, equation~(\ref{eq:matching_optimal}) is equivalent to
\begin{equation}
\frac{\cs}{\Gamma^{2}} \propto U_{0}.
\label{eq:clue}
\end{equation}
Relation~(\ref{eq:clue}) connects the morphological properties of the soft coral ($\Gamma$ and $\cs$) on the left-hand side with the ambient water speed ($U_{0}$) on the right-hand side. This led us to ask the following question: if the soft coral happens to tune its morphology to cope with the local predominant currents, which property should be adjusted? A simple reasoning based on the relation~(\ref{eq:clue}) reveals that in deep waters where the flow is globally calm (small $U_{0}$), soft corals would need thin branches (slender phenotype, large $\Gamma$) and a soft skeleton (small $\cs$). Conversely, in shallow waters where the flow is turbid (high $U_{0}$), they would need thick branches (stout, bushy phenotype, small $\Gamma$) and a stiff skeleton (large $\cs$). Physiological compromise should limit the extreme values of $\cs/ \Gamma^2$. 
A soft coral with both thin branches and a soft skeleton would be too flimsy and pushed down by the flow, which would make it unable to filter or capture any particle. It would also put it at risk of predators. Conversely, a soft coral with both thick and stiff branches would require too much energy to maintain a proper metabolism and a large strain to flex.

\citet{jeyasuria_mechanical_1987} measured the Young's modulus of the skeleton for several soft corals, and found that deep-water species are stiffer than shallow-water species. Moreover, the morphological comparison between bipinnate sea plumes in different habitats carried out by \citet{sanchez_phenotypic_2007} indicated that deep-water corals have larger aspect ratios than their shallow-water counterparts. Considering these ecological observations, we conjecture that tuning the aspect ratio $\Gamma$ is the solution that soft corals rely on to maximise particle interception. This fact can also be intuitively inferred if we notice that the aspect ratio is squared in equation~(\ref{eq:clue}).

\section{Conclusion and future work}
In this paper, we inquire about the observed high-frequency vibrations of the branches of the soft coral \textit{A.~bipinnata}.
First we diagnose the origin of vibration. From numerical and qualitative arguments, we find that VIVs are the most plausible cause of these fast dynamics. Then we investigate the rate of particle interception. We model the coral branch as a cylinder of circular cross-section. Instead of imposing a vibration amplitude or frequency, we allow the cylinder to be free to oscillate under VIVs in both the transverse and stream-wise directions. Furthermore, we assimilate food particles to spheres and subject them to drag, pressure load and added mass force. The simulation of their trajectories and the calculation of the capture rate show that the vibrating cylinder, at lock-in, can intercept up to 40\% more particles than a fixed cylinder. These simulations also reveal, conversely, the existence of a range of reduced velocities after lock-in where the vibrations are detrimental for capture. For this reason, we cautiously avoid affirming that vibrations are either totally beneficial or totally detrimental, and bring back the evaluation of the interception efficiency by referring to the regime of reduced velocities instead.

In the present simulations, we vary the Reynolds numbers and particle sizes. We consider a constant mass ratio of the cylinder $M = 1$ and neglect structural damping $\zeta = 0$. Because increasing the product $M\zeta$ shortens the amplitude of vibration $\Ymax$ \citep{khalak_motions_1999}, we expect from the function $\delta(\Ymax)$ in figure~\ref{fig:Ymax_Xmax_vs_Ur}$(a)$ that the cylinder would lose in particle capture efficiency. Additionally, we consider particles having the same density ratio, but that are twice as heavy as water $\rhoLess = 2$. For a given Reynolds number, the particle density affects the capture efficiency depending on its size. \citet{espinosa-gayosso_density-ratio_2015} showed that for $\Rey \sim 100$, a weakly buoyant particle slightly lighter than water ($\rhoLess = 0.9$) would never beat the efficiency scored by a heavy sediment-type particle ($\rhoLess = 2.6$) if it is big. They also showed that, conversely, the discrepancy between a weakly buoyant particle and a sediment-type particle disappears as long as the particle is small. It would be beneficial to carry out a parametric study and highlight the dependency of the capture efficiency on the density between these two extremes.

\textit{A. bipinnata} possess a three-dimensional tree-like morphology, so the two-dimensional dynamics of our spring-mounted rigid cylinder neglects three-dimensional features that would be observed in nature. It should be interesting for future studies to consider the global three-dimensional multi-modal dynamics of the entire coral colony coupled with the oscillating shear flow for a better representation of the ocean floor. The shape of polyps and their distribution along the coral are also worth examining as they could locally influence the flow streamlines. These suggestions could be implemented either experimentally, as \citet{rodriguez_scaling_2008, rodriguez_multimodal_2012} and \citet{der_loughian_measuring_2014} carried out for plants and trees, or numerically by resolving the 3-D flow on the deforming structure or using a reduced-order model approach.

In addition, our model of particle advection is more appropriate for passive particles. In reality, some particles are motile and might dodge
and escape capture. Although their propulsion force becomes useless when the water current is important, it is worth taking it into account and determine the flow regimes where the motility influences the capture.
Furthermore, instead of the solid contact criterion, a more realistic capture condition may consider some capture zones around polyps, with areas representing the tentacles reach, and each of them having a probability of a successful catch. Another improvement of the capture criterion might be to assign a retention duration to particles directly hitting the cylinder edge to reflect the retaining action of the mucus.

The results of this paper, which are mechanical in essence, may shed light on soft coral research and promote the connection between fluid-structure interaction and invertebrate biology.
In this context, a fruitful avenue would be to see if the relation (\ref{eq:clue}) echoes real data collected during in-field expeditions or \textit{in situ} experiments on \textit{A.~bipinnata} and other colonies. Transplanting soft corals in different habitats and investigating the variation of their natural frequencies and modes in water could also be helpful to understand their biological plasticity and distribution in oceans.

In the end, it is worth reflecting on how a flow-induced instability can be turned into a strategy. At a time when VIVs represent a major threat to offshore energy production, the soft coral paradigm unveils an advantageous side of this instability that can spark novel ideas in biomimetics. Soft corals may, for instance, bioinspire engineers to design innovative energy harvesters at sea, and seek -- instead of suppress -- VIVs as a principal supplier to harness clean and renewable energy.\\

\noindent{\bf Supplementary data\bf{.}} Dynamical equations and numerical methods used in the elastic rod simulations in section \ref{sec:origin}, as well as a verification of the particle advection code in section \ref{sec:advection}, are available at https://doi.org/10.1017/jfm.2021.252. \\

\noindent{\bf Acknowledgements\bf{.}} We would like to thank Camille Soenen for performing measurements on living soft corals and extracting their mechanical properties. \\

\noindent{\bf Funding\bf{.}} The authors acknowledge the financial support from Discovery Grants Nos. RGPIN-2019-07072, RGPIN-2019-05335, as well as from the Simulation-Based Engineering Science (SBES) program through the CREATE grant of the National Science and Engineering Research Council of Canada (NSERC).\\

\noindent{\bf Declaration of Interests\bf{.}} The authors report no conflict of interest. \\

\noindent{\bf Author ORCID\bf{.}} M. Boudina, https://orcid.org/0000-0002-4908-4589; F. P. Gosselin, https://orcid.org/0000-0003-0639-7419; S. Étienne, https://orcid.org/0000-0003-2813-0061.\\

\appendix

\section{Refuting galloping as a cause of soft coral vibrations}
\label{app:galloping}

\begin{figure}
\centerline{
\includegraphics[height=0.275\textheight]{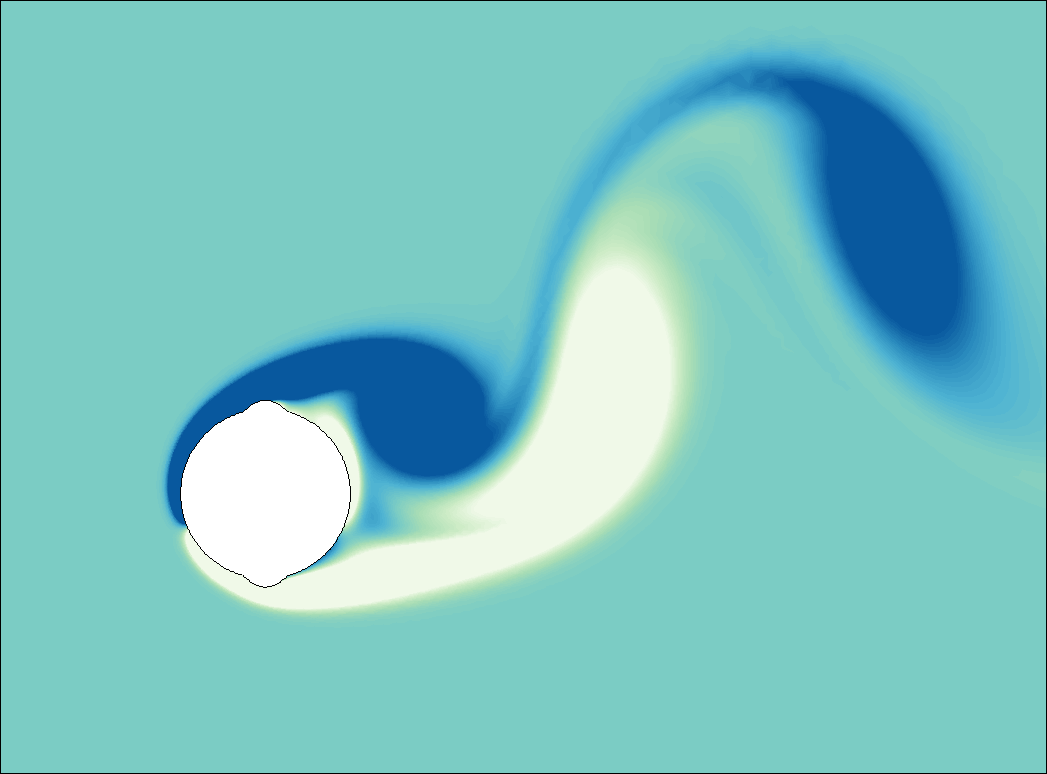}
\includegraphics[height=0.275\textheight]{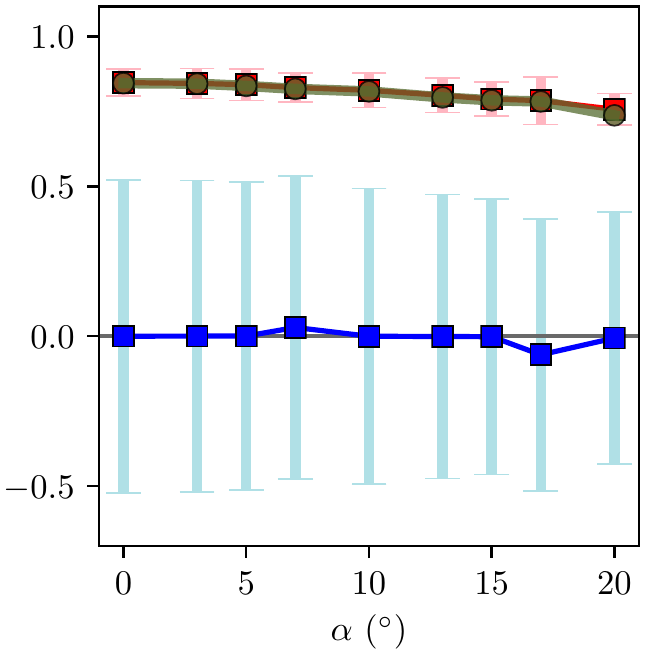}
}
\caption{Left: vorticity contour plot of a flow around an idealised soft coral cross-section at the angle of attack 20$^{\circ}$ and Reynolds number $\Rey = 200$. The cross-section is a circle and polyps are the two diametrically opposed arcs of circle. Their size is 10\% of the circle diameter. Right: mean values of the fluctuating drag and lift coefficients, $\Cd$ (\textcolor{red}{$\fullsquare$}) and $\Cl$ (\textcolor{blue}{$\fullsquare$}). The bars refer to the maximum and minimum values of the variables. The mean value of $\mathrm{d}\Cl/\mathrm{d}\alpha + \Cd$~(\textcolor{darkolivegreen}{$\fullcirc$}) is calculated from cubic spline interpolation of the data points.}
\label{fig:G-denH}
\end{figure}

The dried bipinnate sea plume in figure \ref{fig:branch} has protuberances along its branches, which are not perfect circular cylinders. To determine whether this geometrical perturbation is sufficiently small to render the branch safe from galloping, we considered an idealised cross-section of \textit{A.~bipinnata}, as shown in figure \ref{fig:G-denH}, left. We modelled protuberances as diametrically opposed arcs of a circle having a size of 10\% of the cylinder diameter, to conform to the photography in figure \ref{fig:branch} and data provided by \citet{bayer_shallow-water_1961}. We simulated flows of different angles of attack $\alpha$ around this geometry using the same in-house flow solver \citep{etienne_perspective_2009} described in section \ref{subsec:flow}, then evaluated the Glauert-den Hartog criterion. The latter states that a sufficient condition for galloping to arise is
\begin{equation}
\totalDeriv{\Cl}{\alpha} + \Cd < 0,
\label{eq:G-denH}
\end{equation}
where $\Cl$ and $\Cd$ are the lift and drag coefficients. From a video of the vibrating coral \citep{youtube_caribbean_2013}, and again using the software \textsc{imageJ}, we estimated that the transverse displacement of the branch does not exceed~$\Ymax/D \sim~0.5$. The angle of attack to which the coral cross-section would be exposed is then, at most,~$\tan^{-1}(2\upi\fn A/U_{0}) \sim~24^{\circ}$. Therefore, we chose a representative range of $\alpha$ between 0$^{\circ}$ and 20$^{\circ}$. Figure \ref{fig:G-denH}, right, shows the variation of the fluid-dynamic coefficients and the Glauert-den Hartog criterion. The quantity $\rmdee \Cl / \rmdee\alpha + \Cd$ is always close to 1 and has no tendency to change sign in the range of $\alpha$ considered, which ensures that galloping cannot be a cause of the high frequency motion of the soft coral branches.

%% file: msc_paper_arxiv.bbl
\begin{thebibliography}{79}%
\makeatletter
\providecommand \@ifxundefined [1]{%
 \@ifx{#1\undefined}
}%
\providecommand \@ifnum [1]{%
 \ifnum #1\expandafter \@firstoftwo
 \else \expandafter \@secondoftwo
 \fi
}%
\providecommand \@ifx [1]{%
 \ifx #1\expandafter \@firstoftwo
 \else \expandafter \@secondoftwo
 \fi
}%
\providecommand \natexlab [1]{#1}%
\providecommand \enquote  [1]{``#1''}%
\providecommand \bibnamefont  [1]{#1}%
\providecommand \bibfnamefont [1]{#1}%
\providecommand \citenamefont [1]{#1}%
\providecommand \href@noop [0]{\@secondoftwo}%
\providecommand \href [0]{\begingroup \@sanitize@url \@href}%
\providecommand \@href[1]{\@@startlink{#1}\@@href}%
\providecommand \@@href[1]{\endgroup#1\@@endlink}%
\providecommand \@sanitize@url [0]{\catcode `\\12\catcode `\$12\catcode
  `\&12\catcode `\#12\catcode `\^12\catcode `\_12\catcode `\%12\relax}%
\providecommand \@@startlink[1]{}%
\providecommand \@@endlink[0]{}%
\providecommand \url  [0]{\begingroup\@sanitize@url \@url }%
\providecommand \@url [1]{\endgroup\@href {#1}{\urlprefix }}%
\providecommand \urlprefix  [0]{URL }%
\providecommand \Eprint [0]{\href }%
\providecommand \doibase [0]{http://dx.doi.org/}%
\providecommand \selectlanguage [0]{\@gobble}%
\providecommand \bibinfo  [0]{\@secondoftwo}%
\providecommand \bibfield  [0]{\@secondoftwo}%
\providecommand \translation [1]{[#1]}%
\providecommand \BibitemOpen [0]{}%
\providecommand \bibitemStop [0]{}%
\providecommand \bibitemNoStop [0]{.\EOS\space}%
\providecommand \EOS [0]{\spacefactor3000\relax}%
\providecommand \BibitemShut  [1]{\csname bibitem#1\endcsname}%
\let\auto@bib@innerbib\@empty
\bibitem [{\citenamefont {Ribes}\ \emph {et~al.}(1998)\citenamefont {Ribes},
  \citenamefont {Coma},\ and\ \citenamefont {Gili}}]{ribes_heterotrophic_1998}%
  \BibitemOpen
  \bibfield  {author} {\bibinfo {author} {\bibfnamefont {M.}~\bibnamefont
  {Ribes}}, \bibinfo {author} {\bibfnamefont {R.}~\bibnamefont {Coma}}, \ and\
  \bibinfo {author} {\bibfnamefont {J.-M.}\ \bibnamefont {Gili}},\ }\href
  {\doibase 10.4319/lo.1998.43.6.1170} {\bibfield  {journal} {\bibinfo
  {journal} {Limnology and Oceanography}\ }\textbf {\bibinfo {volume} {43}},\
  \bibinfo {pages} {1170} (\bibinfo {year} {1998})},\ \bibinfo {note}
  {publisher: John Wiley \& Sons, Ltd}\BibitemShut {NoStop}%
\bibitem [{\citenamefont {Wainwright}\ and\ \citenamefont
  {Dillon}(1969)}]{wainwright_orientation_1969}%
  \BibitemOpen
  \bibfield  {author} {\bibinfo {author} {\bibfnamefont {S.~A.}\ \bibnamefont
  {Wainwright}}\ and\ \bibinfo {author} {\bibfnamefont {J.~R.}\ \bibnamefont
  {Dillon}},\ }\href {\doibase 10.2307/1539674} {\bibfield  {journal} {\bibinfo
   {journal} {Biol Bull}\ }\textbf {\bibinfo {volume} {136}} (\bibinfo {year}
  {1969}),\ 10.2307/1539674}\BibitemShut {NoStop}%
\bibitem [{\citenamefont {Williams}\ and\ \citenamefont
  {Chen}(2012)}]{williams_resurrection_2012}%
  \BibitemOpen
  \bibfield  {author} {\bibinfo {author} {\bibfnamefont {G.~C.}\ \bibnamefont
  {Williams}}\ and\ \bibinfo {author} {\bibfnamefont {J.-Y.}\ \bibnamefont
  {Chen}},\ }\href@noop {} {\bibfield  {journal} {\bibinfo  {journal}
  {Zootaxa}\ }\textbf {\bibinfo {volume} {3505}},\ \bibinfo {pages} {39}
  (\bibinfo {year} {2012})}\BibitemShut {NoStop}%
\bibitem [{\citenamefont {Gosselin}(2019)}]{gosselin_mechanics_2019}%
  \BibitemOpen
  \bibfield  {author} {\bibinfo {author} {\bibfnamefont {F.~P.}\ \bibnamefont
  {Gosselin}},\ }\href {\doibase 10.1093/jxb/erz288} {\bibfield  {journal}
  {\bibinfo  {journal} {Journal of Experimental Botany}\ }\textbf {\bibinfo
  {volume} {70}},\ \bibinfo {pages} {3533} (\bibinfo {year}
  {2019})}\BibitemShut {NoStop}%
\bibitem [{\citenamefont {Vogel}(1984)}]{vogel_drag_1984}%
  \BibitemOpen
  \bibfield  {author} {\bibinfo {author} {\bibfnamefont {S.}~\bibnamefont
  {Vogel}},\ }\href@noop {} {\bibfield  {journal} {\bibinfo  {journal}
  {American Zoologist}\ }\textbf {\bibinfo {volume} {24}},\ \bibinfo {pages}
  {37} (\bibinfo {year} {1984})}\BibitemShut {NoStop}%
\bibitem [{\citenamefont {de~Langre}(2008)}]{de_langre_effects_2008}%
  \BibitemOpen
  \bibfield  {author} {\bibinfo {author} {\bibfnamefont {E.}~\bibnamefont
  {de~Langre}},\ }\href {\doibase 10.1146/annurev.fluid.40.111406.102135}
  {\bibfield  {journal} {\bibinfo  {journal} {Annual Review of Fluid
  Mechanics}\ }\textbf {\bibinfo {volume} {40}},\ \bibinfo {pages} {141}
  (\bibinfo {year} {2008})}\BibitemShut {NoStop}%
\bibitem [{\citenamefont {Jeyasuria}\ and\ \citenamefont
  {Lewis}(1987)}]{jeyasuria_mechanical_1987}%
  \BibitemOpen
  \bibfield  {author} {\bibinfo {author} {\bibfnamefont {P.}~\bibnamefont
  {Jeyasuria}}\ and\ \bibinfo {author} {\bibfnamefont {J.}~\bibnamefont
  {Lewis}},\ }\href@noop {} {\bibfield  {journal} {\bibinfo  {journal} {Coral
  reefs}\ }\textbf {\bibinfo {volume} {5}},\ \bibinfo {pages} {213} (\bibinfo
  {year} {1987})}\BibitemShut {NoStop}%
\bibitem [{\citenamefont {Sánchez}\ \emph {et~al.}(2007)\citenamefont
  {Sánchez}, \citenamefont {Aguilar}, \citenamefont {Dorado},\ and\
  \citenamefont {Manrique}}]{sanchez_phenotypic_2007}%
  \BibitemOpen
  \bibfield  {author} {\bibinfo {author} {\bibfnamefont {J.~A.}\ \bibnamefont
  {Sánchez}}, \bibinfo {author} {\bibfnamefont {C.}~\bibnamefont {Aguilar}},
  \bibinfo {author} {\bibfnamefont {D.}~\bibnamefont {Dorado}}, \ and\ \bibinfo
  {author} {\bibfnamefont {N.}~\bibnamefont {Manrique}},\ }\href {\doibase
  10.1186/1471-2148-7-122} {\bibfield  {journal} {\bibinfo  {journal} {BMC
  Evolutionary Biology}\ }\textbf {\bibinfo {volume} {7}},\ \bibinfo {pages}
  {122} (\bibinfo {year} {2007})}\BibitemShut {NoStop}%
\bibitem [{\citenamefont {Leclercq}\ and\ \citenamefont {{de
  }Langre}(2018)}]{leclercq_reconfiguration_2018}%
  \BibitemOpen
  \bibfield  {author} {\bibinfo {author} {\bibfnamefont {T.}~\bibnamefont
  {Leclercq}}\ and\ \bibinfo {author} {\bibfnamefont {E.}~\bibnamefont {{de
  }Langre}},\ }\href {\doibase 10.1017/jfm.2017.910} {\bibfield  {journal}
  {\bibinfo  {journal} {Journal of Fluid Mechanics}\ }\textbf {\bibinfo
  {volume} {838}},\ \bibinfo {pages} {606} (\bibinfo {year}
  {2018})}\BibitemShut {NoStop}%
\bibitem [{\citenamefont {Lei}\ and\ \citenamefont
  {Nepf}(2019)}]{lei_blade_2019}%
  \BibitemOpen
  \bibfield  {author} {\bibinfo {author} {\bibfnamefont {J.}~\bibnamefont
  {Lei}}\ and\ \bibinfo {author} {\bibfnamefont {H.}~\bibnamefont {Nepf}},\
  }\href {\doibase 10.1016/j.jfluidstructs.2019.03.020} {\bibfield  {journal}
  {\bibinfo  {journal} {Journal of Fluids and Structures}\ }\textbf {\bibinfo
  {volume} {87}},\ \bibinfo {pages} {137} (\bibinfo {year} {2019})}\BibitemShut
  {NoStop}%
\bibitem [{\citenamefont {Rodriguez}\ \emph {et~al.}(2008)\citenamefont
  {Rodriguez}, \citenamefont {{de Langre}},\ and\ \citenamefont
  {Moulia}}]{rodriguez_scaling_2008}%
  \BibitemOpen
  \bibfield  {author} {\bibinfo {author} {\bibfnamefont {M.}~\bibnamefont
  {Rodriguez}}, \bibinfo {author} {\bibfnamefont {E.}~\bibnamefont {{de
  Langre}}}, \ and\ \bibinfo {author} {\bibfnamefont {B.}~\bibnamefont
  {Moulia}},\ }\href {\doibase 10.3732/ajb.0800161} {\bibfield  {journal}
  {\bibinfo  {journal} {American Journal of Botany}\ }\textbf {\bibinfo
  {volume} {95}},\ \bibinfo {pages} {1523} (\bibinfo {year}
  {2008})}\BibitemShut {NoStop}%
\bibitem [{\citenamefont {Rodriguez}\ \emph {et~al.}(2012)\citenamefont
  {Rodriguez}, \citenamefont {Ploquin}, \citenamefont {Moulia},\ and\
  \citenamefont {{de }Langre}}]{rodriguez_multimodal_2012}%
  \BibitemOpen
  \bibfield  {author} {\bibinfo {author} {\bibfnamefont {M.}~\bibnamefont
  {Rodriguez}}, \bibinfo {author} {\bibfnamefont {S.}~\bibnamefont {Ploquin}},
  \bibinfo {author} {\bibfnamefont {B.}~\bibnamefont {Moulia}}, \ and\ \bibinfo
  {author} {\bibfnamefont {E.}~\bibnamefont {{de }Langre}},\ }\href
  {https://manufacturingscience.asmedigitalcollection.asme.org/data/Journals/JAMCAV/26820/044505_1.pdf}
  {\bibfield  {journal} {\bibinfo  {journal} {Journal of Applied Mechanics}\
  }\textbf {\bibinfo {volume} {79}},\ \bibinfo {pages} {044505} (\bibinfo
  {year} {2012})}\BibitemShut {NoStop}%
\bibitem [{\citenamefont {Der~Loughian}\ \emph {et~al.}(2014)\citenamefont
  {Der~Loughian}, \citenamefont {Tadrist}, \citenamefont {Allain},
  \citenamefont {Diener}, \citenamefont {Moulia},\ and\ \citenamefont {{de
  Langre}}}]{der_loughian_measuring_2014}%
  \BibitemOpen
  \bibfield  {author} {\bibinfo {author} {\bibfnamefont {C.}~\bibnamefont
  {Der~Loughian}}, \bibinfo {author} {\bibfnamefont {L.}~\bibnamefont
  {Tadrist}}, \bibinfo {author} {\bibfnamefont {J.-M.}\ \bibnamefont {Allain}},
  \bibinfo {author} {\bibfnamefont {J.}~\bibnamefont {Diener}}, \bibinfo
  {author} {\bibfnamefont {B.}~\bibnamefont {Moulia}}, \ and\ \bibinfo {author}
  {\bibfnamefont {E.}~\bibnamefont {{de Langre}}},\ }\href {\doibase
  10.1016/j.crme.2013.10.010} {\bibfield  {journal} {\bibinfo  {journal}
  {Comptes Rendus Mécanique}\ }\textbf {\bibinfo {volume} {342}},\ \bibinfo
  {pages} {1} (\bibinfo {year} {2014})}\BibitemShut {NoStop}%
\bibitem [{\citenamefont {Williamson}\ and\ \citenamefont
  {Govardhan}(2004)}]{williamson_vortex-induced_2004}%
  \BibitemOpen
  \bibfield  {author} {\bibinfo {author} {\bibfnamefont {C.}~\bibnamefont
  {Williamson}}\ and\ \bibinfo {author} {\bibfnamefont {R.}~\bibnamefont
  {Govardhan}},\ }\href {\doibase 10.1146/annurev.fluid.36.050802.122128}
  {\bibfield  {journal} {\bibinfo  {journal} {Annual Review of Fluid
  Mechanics}\ }\textbf {\bibinfo {volume} {36}},\ \bibinfo {pages} {413}
  (\bibinfo {year} {2004})}\BibitemShut {NoStop}%
\bibitem [{\citenamefont {Sarpkaya}(2010)}]{sarpkaya_wave_2010}%
  \BibitemOpen
  \bibfield  {author} {\bibinfo {author} {\bibfnamefont {T.}~\bibnamefont
  {Sarpkaya}},\ }\href@noop {} {{\emph {\bibinfo {title}
  {Wave {Forces} on {Offshore} {Structures}}}}}\ (\bibinfo  {publisher}
  {Cambridge University Press},\ \bibinfo {year} {2010})\BibitemShut {NoStop}%
\bibitem [{\citenamefont {Freds\o{}e}\ and\ \citenamefont
  {Sumer}(2006)}]{fredsoe_hydrodynamics_2006}%
  \BibitemOpen
  \bibfield  {author} {\bibinfo {author} {\bibfnamefont {J.}~\bibnamefont
  {Freds\o{}e}}\ and\ \bibinfo {author} {\bibfnamefont {M.~B.}\ \bibnamefont
  {Sumer}},\ }\href@noop {} {{\emph {\bibinfo {title}
  {Hydrodynamics {Around} {Cylindrical} {Structures} ({Revised} {Edition})}}}}\
  (\bibinfo  {publisher} {World Scientific},\ \bibinfo {year}
  {2006})\BibitemShut {NoStop}%
\bibitem [{\citenamefont {Etienne}\ and\ \citenamefont
  {Pelletier}(2012)}]{etienne_low_2012}%
  \BibitemOpen
  \bibfield  {author} {\bibinfo {author} {\bibfnamefont {S.}~\bibnamefont
  {Etienne}}\ and\ \bibinfo {author} {\bibfnamefont {D.}~\bibnamefont
  {Pelletier}},\ }\href {\doibase 10.1016/j.jfluidstructs.2012.02.006}
  {\bibfield  {journal} {\bibinfo  {journal} {Journal of Fluids and
  Structures}\ }\textbf {\bibinfo {volume} {31}},\ \bibinfo {pages} {18}
  (\bibinfo {year} {2012})}\BibitemShut {NoStop}%
\bibitem [{\citenamefont {Bishop}\ and\ \citenamefont
  {Hassan}(1964)}]{bishop_lift_1964}%
  \BibitemOpen
  \bibfield  {author} {\bibinfo {author} {\bibfnamefont {R.~E.~D.}\
  \bibnamefont {Bishop}}\ and\ \bibinfo {author} {\bibfnamefont {A.~Y.}\
  \bibnamefont {Hassan}},\ }\href {\doibase 10.1098/rspa.1964.0005} {\bibfield
  {journal} {\bibinfo  {journal} {Proc. R. Soc. Lond. A}\ }\textbf {\bibinfo
  {volume} {277}},\ \bibinfo {pages} {51} (\bibinfo {year} {1964})}\BibitemShut
  {NoStop}%
\bibitem [{\citenamefont {Chaplin}\ \emph {et~al.}(2005)\citenamefont
  {Chaplin}, \citenamefont {Bearman}, \citenamefont {Huera~Huarte},\ and\
  \citenamefont {Pattenden}}]{chaplin_laboratory_2005}%
  \BibitemOpen
  \bibfield  {author} {\bibinfo {author} {\bibfnamefont {J.~R.}\ \bibnamefont
  {Chaplin}}, \bibinfo {author} {\bibfnamefont {P.~W.}\ \bibnamefont
  {Bearman}}, \bibinfo {author} {\bibfnamefont {F.~J.}\ \bibnamefont
  {Huera~Huarte}}, \ and\ \bibinfo {author} {\bibfnamefont {R.~J.}\
  \bibnamefont {Pattenden}},\ }\href {\doibase
  10.1016/j.jfluidstructs.2005.04.010} {\bibfield  {journal} {\bibinfo
  {journal} {Journal of Fluids and Structures}\ }\bibinfo {series}
  {Fluid-{Structure} and {Flow}-{Acoustic} {Interactions} involving {Bluff}
  {Bodies}},\ \textbf {\bibinfo {volume} {21}},\ \bibinfo {pages} {3} (\bibinfo
  {year} {2005})}\BibitemShut {NoStop}%
\bibitem [{\citenamefont {Newman}\ and\ \citenamefont
  {Karniadakis}(1997)}]{newman_direct_1997}%
  \BibitemOpen
  \bibfield  {author} {\bibinfo {author} {\bibfnamefont {D.~J.}\ \bibnamefont
  {Newman}}\ and\ \bibinfo {author} {\bibfnamefont {G.~E.}\ \bibnamefont
  {Karniadakis}},\ }\href {\doibase 10.1017/S002211209700582X} {\bibfield
  {journal} {\bibinfo  {journal} {Journal of Fluid Mechanics}\ }\textbf
  {\bibinfo {volume} {344}},\ \bibinfo {pages} {95} (\bibinfo {year}
  {1997})}\BibitemShut {NoStop}%
\bibitem [{\citenamefont {Facchinetti}\ \emph
  {et~al.}(2004{\natexlab{a}})\citenamefont {Facchinetti}, \citenamefont {{de
  Langre}},\ and\ \citenamefont {Biolley}}]{facchinetti_vortex-induced_2004}%
  \BibitemOpen
  \bibfield  {author} {\bibinfo {author} {\bibfnamefont {M.~L.}\ \bibnamefont
  {Facchinetti}}, \bibinfo {author} {\bibfnamefont {E.}~\bibnamefont {{de
  Langre}}}, \ and\ \bibinfo {author} {\bibfnamefont {F.}~\bibnamefont
  {Biolley}},\ }\href {\doibase 10.1016/j.euromechflu.2003.04.004} {\bibfield
  {journal} {\bibinfo  {journal} {European Journal of Mechanics - B/Fluids}\
  }\bibinfo {series} {Bluff {Body} {Wakes} and {Vortex}-{Induced}
  {Vibrations}},\ \textbf {\bibinfo {volume} {23}},\ \bibinfo {pages} {199}
  (\bibinfo {year} {2004}{\natexlab{a}})}\BibitemShut {NoStop}%
\bibitem [{\citenamefont {Evangelinos}\ and\ \citenamefont
  {Karniadakis}(1999)}]{evangelinos_dynamics_1999}%
  \BibitemOpen
  \bibfield  {author} {\bibinfo {author} {\bibfnamefont {C.}~\bibnamefont
  {Evangelinos}}\ and\ \bibinfo {author} {\bibfnamefont {G.~E.}\ \bibnamefont
  {Karniadakis}},\ }\href {\doibase 10.1017/S0022112099006606} {\bibfield
  {journal} {\bibinfo  {journal} {Journal of Fluid Mechanics}\ }\textbf
  {\bibinfo {volume} {400}},\ \bibinfo {pages} {91} (\bibinfo {year} {1999})},\
  \bibinfo {note} {publisher: Cambridge University Press}\BibitemShut {NoStop}%
\bibitem [{\citenamefont {Lucor}\ \emph {et~al.}(2006)\citenamefont {Lucor},
  \citenamefont {Mukundan},\ and\ \citenamefont
  {Triantafyllou}}]{lucor_riser_2006}%
  \BibitemOpen
  \bibfield  {author} {\bibinfo {author} {\bibfnamefont {D.}~\bibnamefont
  {Lucor}}, \bibinfo {author} {\bibfnamefont {H.}~\bibnamefont {Mukundan}}, \
  and\ \bibinfo {author} {\bibfnamefont {M.~S.}\ \bibnamefont
  {Triantafyllou}},\ }\href {\doibase 10.1016/j.jfluidstructs.2006.04.006}
  {\bibfield  {journal} {\bibinfo  {journal} {Journal of Fluids and
  Structures}\ }\bibinfo {series} {"{Bluff} {Body} {Wakes} and
  {Vortex}-{Induced} {Vibrations} ({BBVIV}-4)},\ \textbf {\bibinfo {volume}
  {22}},\ \bibinfo {pages} {905} (\bibinfo {year} {2006})}\BibitemShut
  {NoStop}%
\bibitem [{\citenamefont {Violette}\ \emph {et~al.}(2007)\citenamefont
  {Violette}, \citenamefont {{de }Langre},\ and\ \citenamefont
  {Szydlowski}}]{violette_computation_2007}%
  \BibitemOpen
  \bibfield  {author} {\bibinfo {author} {\bibfnamefont {R.}~\bibnamefont
  {Violette}}, \bibinfo {author} {\bibfnamefont {E.}~\bibnamefont {{de
  }Langre}}, \ and\ \bibinfo {author} {\bibfnamefont {J.}~\bibnamefont
  {Szydlowski}},\ }\href {\doibase 10.1016/j.compstruc.2006.08.005} {\bibfield
  {journal} {\bibinfo  {journal} {Computers \& Structures}\ }\bibinfo {series}
  {Fourth {MIT} {Conference} on {Computational} {Fluid} and {Solid}
  {Mechanics}},\ \textbf {\bibinfo {volume} {85}},\ \bibinfo {pages} {1134}
  (\bibinfo {year} {2007})}\BibitemShut {NoStop}%
\bibitem [{\citenamefont {Kim}\ \emph {et~al.}(2019)\citenamefont {Kim},
  \citenamefont {Park}, \citenamefont {Gruszewski}, \citenamefont {Schmale},\
  and\ \citenamefont {Jung}}]{kim_vortex-induced_2019}%
  \BibitemOpen
  \bibfield  {author} {\bibinfo {author} {\bibfnamefont {S.}~\bibnamefont
  {Kim}}, \bibinfo {author} {\bibfnamefont {H.}~\bibnamefont {Park}}, \bibinfo
  {author} {\bibfnamefont {H.~A.}\ \bibnamefont {Gruszewski}}, \bibinfo
  {author} {\bibfnamefont {D.~G.}\ \bibnamefont {Schmale}}, \ and\ \bibinfo
  {author} {\bibfnamefont {S.}~\bibnamefont {Jung}},\ }\href {\doibase
  10.1073/pnas.1820318116} {\bibfield  {journal} {\bibinfo  {journal}
  {Proceedings of the National Academy of Sciences}\ }\textbf {\bibinfo
  {volume} {116}},\ \bibinfo {pages} {4917} (\bibinfo {year} {2019})},\
  \bibinfo {note} {publisher: National Academy of Sciences Section: Biological
  Sciences}\BibitemShut {NoStop}%
\bibitem [{\citenamefont {Gilpin}\ \emph {et~al.}(2017)\citenamefont {Gilpin},
  \citenamefont {Prakash},\ and\ \citenamefont {Prakash}}]{gilpin2017}%
  \BibitemOpen
  \bibfield  {author} {\bibinfo {author} {\bibfnamefont {W.}~\bibnamefont
  {Gilpin}}, \bibinfo {author} {\bibfnamefont {V.~N.}\ \bibnamefont {Prakash}},
  \ and\ \bibinfo {author} {\bibfnamefont {M.}~\bibnamefont {Prakash}},\ }\href
  {\doibase 10.1038/nphys3981} {\bibfield  {journal} {\bibinfo  {journal}
  {Nature Physics}\ }\textbf {\bibinfo {volume} {13}},\ \bibinfo {pages} {380}
  (\bibinfo {year} {2017})}\BibitemShut {NoStop}%
\bibitem [{\citenamefont {Chance}\ and\ \citenamefont
  {Craig}(1986)}]{chance_hydrodynamics_1986}%
  \BibitemOpen
  \bibfield  {author} {\bibinfo {author} {\bibfnamefont {M.~M.}\ \bibnamefont
  {Chance}}\ and\ \bibinfo {author} {\bibfnamefont {D.~A.}\ \bibnamefont
  {Craig}},\ }\href {\doibase 10.1139/z86-193} {\bibfield  {journal} {\bibinfo
  {journal} {Canadian Journal of Zoology}\ }\textbf {\bibinfo {volume} {64}},\
  \bibinfo {pages} {1295} (\bibinfo {year} {1986})},\ \bibinfo {note}
  {publisher: NRC Research Press}\BibitemShut {NoStop}%
\bibitem [{\citenamefont {Widahl}(1992)}]{widahl_flow_1992}%
  \BibitemOpen
  \bibfield  {author} {\bibinfo {author} {\bibfnamefont {L.-E.}\ \bibnamefont
  {Widahl}},\ }\href {\doibase 10.1093/aesa/85.1.91} {\bibfield  {journal}
  {\bibinfo  {journal} {Annals of the Entomological Society of America}\
  }\textbf {\bibinfo {volume} {85}},\ \bibinfo {pages} {91} (\bibinfo {year}
  {1992})},\ \bibinfo {note} {publisher: Oxford Academic}\BibitemShut {NoStop}%
\bibitem [{\citenamefont {Fabricius}(2011)}]{fabricius_octocorallia_2011}%
  \BibitemOpen
  \bibfield  {author} {\bibinfo {author} {\bibfnamefont {K.~E.}\ \bibnamefont
  {Fabricius}},\ }in\ \href@noop {} {\emph {\bibinfo {booktitle} {Encyclopedia
  of {Modern} {Coral} {Reefs}}}}\ (\bibinfo  {publisher} {Springer Science \&
  Business Media},\ \bibinfo {year} {2011})\ pp.\ \bibinfo {pages}
  {740--745}\BibitemShut {NoStop}%
\bibitem [{\citenamefont {Veron}(2011)}]{veron_corals_2011}%
  \BibitemOpen
  \bibfield  {author} {\bibinfo {author} {\bibfnamefont {J.~E.~N.}\
  \bibnamefont {Veron}},\ }in\ \href@noop {} {\emph {\bibinfo {booktitle}
  {Encyclopedia of {Modern} {Coral} {Reefs}}}}\ (\bibinfo  {publisher}
  {Springer Science \& Business Media},\ \bibinfo {year} {2011})\ pp.\ \bibinfo
  {pages} {275--281}\BibitemShut {NoStop}%
\bibitem [{\citenamefont {Shimeta}\ and\ \citenamefont
  {Koehl}(1997)}]{shimeta_mechanisms_1997}%
  \BibitemOpen
  \bibfield  {author} {\bibinfo {author} {\bibfnamefont {J.}~\bibnamefont
  {Shimeta}}\ and\ \bibinfo {author} {\bibfnamefont {M.~A.~R.}\ \bibnamefont
  {Koehl}},\ }\href {\doibase 10.1016/S0022-0981(96)02684-6} {\bibfield
  {journal} {\bibinfo  {journal} {Journal of Experimental Marine Biology and
  Ecology}\ }\textbf {\bibinfo {volume} {209}},\ \bibinfo {pages} {47}
  (\bibinfo {year} {1997})}\BibitemShut {NoStop}%
\bibitem [{\citenamefont {Weber}\ and\ \citenamefont
  {Paddock}(1983)}]{weber_interceptional_1983}%
  \BibitemOpen
  \bibfield  {author} {\bibinfo {author} {\bibfnamefont {M.~E.}\ \bibnamefont
  {Weber}}\ and\ \bibinfo {author} {\bibfnamefont {D.}~\bibnamefont
  {Paddock}},\ }\href {\doibase 10.1016/0021-9797(83)90270-9} {\bibfield
  {journal} {\bibinfo  {journal} {Journal of Colloid and Interface Science}\
  }\textbf {\bibinfo {volume} {94}},\ \bibinfo {pages} {328} (\bibinfo {year}
  {1983})}\BibitemShut {NoStop}%
\bibitem [{\citenamefont {Palmer}\ \emph {et~al.}(2004)\citenamefont {Palmer},
  \citenamefont {Nepf}, \citenamefont {Pettersson},\ and\ \citenamefont
  {Ackerman}}]{palmer_observations_2004}%
  \BibitemOpen
  \bibfield  {author} {\bibinfo {author} {\bibfnamefont {M.~R.}\ \bibnamefont
  {Palmer}}, \bibinfo {author} {\bibfnamefont {H.~M.}\ \bibnamefont {Nepf}},
  \bibinfo {author} {\bibfnamefont {T.~J.~R.}\ \bibnamefont {Pettersson}}, \
  and\ \bibinfo {author} {\bibfnamefont {J.~D.}\ \bibnamefont {Ackerman}},\
  }\href {\doibase 10.4319/lo.2004.49.1.0076} {\bibfield  {journal} {\bibinfo
  {journal} {Limnology and Oceanography}\ }\textbf {\bibinfo {volume} {49}},\
  \bibinfo {pages} {76} (\bibinfo {year} {2004})}\BibitemShut {NoStop}%
\bibitem [{\citenamefont {Haugen}\ and\ \citenamefont
  {Kragset}(2010)}]{haugen_particle_2010}%
  \BibitemOpen
  \bibfield  {author} {\bibinfo {author} {\bibfnamefont {N.~E.~L.}\
  \bibnamefont {Haugen}}\ and\ \bibinfo {author} {\bibfnamefont
  {S.}~\bibnamefont {Kragset}},\ }\href {\doibase 10.1017/S0022112010002946}
  {\bibfield  {journal} {\bibinfo  {journal} {Journal of Fluid Mechanics}\
  }\textbf {\bibinfo {volume} {661}},\ \bibinfo {pages} {239} (\bibinfo {year}
  {2010})}\BibitemShut {NoStop}%
\bibitem [{\citenamefont {Espinosa-Gayosso}\ \emph {et~al.}(2012)\citenamefont
  {Espinosa-Gayosso}, \citenamefont {Ghisalberti}, \citenamefont {Ivey},\ and\
  \citenamefont {Jones}}]{espinosa-gayosso_particle_2012}%
  \BibitemOpen
  \bibfield  {author} {\bibinfo {author} {\bibfnamefont {A.}~\bibnamefont
  {Espinosa-Gayosso}}, \bibinfo {author} {\bibfnamefont {M.}~\bibnamefont
  {Ghisalberti}}, \bibinfo {author} {\bibfnamefont {G.~N.}\ \bibnamefont
  {Ivey}}, \ and\ \bibinfo {author} {\bibfnamefont {N.~L.}\ \bibnamefont
  {Jones}},\ }\href {\doibase 10.1017/jfm.2012.367} {\bibfield  {journal}
  {\bibinfo  {journal} {Journal of Fluid Mechanics}\ }\textbf {\bibinfo
  {volume} {710}},\ \bibinfo {pages} {362} (\bibinfo {year}
  {2012})}\BibitemShut {NoStop}%
\bibitem [{\citenamefont {Espinosa-Gayosso}\ \emph {et~al.}(2013)\citenamefont
  {Espinosa-Gayosso}, \citenamefont {Ghisalberti}, \citenamefont {Ivey},\ and\
  \citenamefont {Jones}}]{espinosa-gayosso_particle_2013}%
  \BibitemOpen
  \bibfield  {author} {\bibinfo {author} {\bibfnamefont {A.}~\bibnamefont
  {Espinosa-Gayosso}}, \bibinfo {author} {\bibfnamefont {M.}~\bibnamefont
  {Ghisalberti}}, \bibinfo {author} {\bibfnamefont {G.~N.}\ \bibnamefont
  {Ivey}}, \ and\ \bibinfo {author} {\bibfnamefont {N.~L.}\ \bibnamefont
  {Jones}},\ }\href {\doibase 10.1017/jfm.2013.407} {\bibfield  {journal}
  {\bibinfo  {journal} {Journal of Fluid Mechanics}\ }\textbf {\bibinfo
  {volume} {733}},\ \bibinfo {pages} {171} (\bibinfo {year}
  {2013})}\BibitemShut {NoStop}%
\bibitem [{\citenamefont {Krick}\ and\ \citenamefont
  {Ackerman}(2015)}]{krick_adding_2015}%
  \BibitemOpen
  \bibfield  {author} {\bibinfo {author} {\bibfnamefont {J.}~\bibnamefont
  {Krick}}\ and\ \bibinfo {author} {\bibfnamefont {J.~D.}\ \bibnamefont
  {Ackerman}},\ }\href {\doibase 10.1016/j.jtbi.2014.12.003} {\bibfield
  {journal} {\bibinfo  {journal} {Journal of Theoretical Biology}\ }\textbf
  {\bibinfo {volume} {368}},\ \bibinfo {pages} {13} (\bibinfo {year}
  {2015})}\BibitemShut {NoStop}%
\bibitem [{\citenamefont {McCombe}\ and\ \citenamefont
  {Ackerman}(2018)}]{mccombe_collector_2018}%
  \BibitemOpen
  \bibfield  {author} {\bibinfo {author} {\bibfnamefont {D.}~\bibnamefont
  {McCombe}}\ and\ \bibinfo {author} {\bibfnamefont {J.~D.}\ \bibnamefont
  {Ackerman}},\ }\href {\doibase 10.1086/697551} {\bibfield  {journal}
  {\bibinfo  {journal} {The American Naturalist}\ }\textbf {\bibinfo {volume}
  {192}},\ \bibinfo {pages} {81} (\bibinfo {year} {2018})}\BibitemShut
  {NoStop}%
\bibitem [{\citenamefont {Niklas}(2015)}]{niklas_biophysical_2015}%
  \BibitemOpen
  \bibfield  {author} {\bibinfo {author} {\bibfnamefont {K.~J.}\ \bibnamefont
  {Niklas}},\ }\href {\doibase 10.1007/s12229-014-9148-9} {\bibfield  {journal}
  {\bibinfo  {journal} {The Botanical Review}\ }\textbf {\bibinfo {volume}
  {81}},\ \bibinfo {pages} {28} (\bibinfo {year} {2015})}\BibitemShut {NoStop}%
\bibitem [{\citenamefont {Inoue}\ \emph {et~al.}(2013)\citenamefont {Inoue},
  \citenamefont {Kayanne}, \citenamefont {Yamamoto},\ and\ \citenamefont
  {Kurihara}}]{inoue_spatial_2013}%
  \BibitemOpen
  \bibfield  {author} {\bibinfo {author} {\bibfnamefont {S.}~\bibnamefont
  {Inoue}}, \bibinfo {author} {\bibfnamefont {H.}~\bibnamefont {Kayanne}},
  \bibinfo {author} {\bibfnamefont {S.}~\bibnamefont {Yamamoto}}, \ and\
  \bibinfo {author} {\bibfnamefont {H.}~\bibnamefont {Kurihara}},\ }\href@noop
  {} {\bibfield  {journal} {\bibinfo  {journal} {Nature Climate Change}\
  }\textbf {\bibinfo {volume} {3}},\ \bibinfo {pages} {683} (\bibinfo {year}
  {2013})}\BibitemShut {NoStop}%
\bibitem [{\citenamefont {Tsounis}\ and\ \citenamefont
  {Edmunds}(2017)}]{tsounis_three_2017}%
  \BibitemOpen
  \bibfield  {author} {\bibinfo {author} {\bibfnamefont {G.}~\bibnamefont
  {Tsounis}}\ and\ \bibinfo {author} {\bibfnamefont {P.~J.}\ \bibnamefont
  {Edmunds}},\ }\href@noop {} {\bibfield  {journal} {\bibinfo  {journal}
  {Ecosphere}\ }\textbf {\bibinfo {volume} {8}},\ \bibinfo {pages} {e01646}
  (\bibinfo {year} {2017})}\BibitemShut {NoStop}%
\bibitem [{\citenamefont {Monismith}(2007)}]{monismith_hydrodynamics_2007}%
  \BibitemOpen
  \bibfield  {author} {\bibinfo {author} {\bibfnamefont {S.~G.}\ \bibnamefont
  {Monismith}},\ }\href@noop {} {\bibfield  {journal} {\bibinfo  {journal}
  {Annual Review of Fluid Mechanics}\ }\textbf {\bibinfo {volume} {39}},\
  \bibinfo {pages} {37} (\bibinfo {year} {2007})}\BibitemShut {NoStop}%
\bibitem [{\citenamefont {YouTube}(2013)}]{youtube_caribbean_2013}%
  \BibitemOpen
  \bibfield  {author} {\bibinfo {author} {\bibnamefont {YouTube}},\ }\href@noop
  {} {\enquote {\bibinfo {title} {Caribbean {Spiny} {Lobster} and a {Bipinnate}
  {Sea} {Plume} coral},}\ } (\bibinfo {year} {2013}),\ \bibinfo {note}
  {https://youtu.be/1mgAyFgYfYw}\BibitemShut {NoStop}%
\bibitem [{\citenamefont {Schindelin}\ \emph {et~al.}(2012)\citenamefont
  {Schindelin}, \citenamefont {Arganda-Carreras}, \citenamefont {Frise},
  \citenamefont {Kaynig}, \citenamefont {Longair}, \citenamefont {Pietzsch},
  \citenamefont {Preibisch}, \citenamefont {Rueden}, \citenamefont {Saalfeld},
  \citenamefont {Schmid}, \citenamefont {Tinevez}, \citenamefont {White},
  \citenamefont {Hartenstein}, \citenamefont {Eliceiri}, \citenamefont
  {Tomancak},\ and\ \citenamefont {Cardona}}]{schindelin_fiji_2012}%
  \BibitemOpen
  \bibfield  {author} {\bibinfo {author} {\bibfnamefont {J.}~\bibnamefont
  {Schindelin}}, \bibinfo {author} {\bibfnamefont {I.}~\bibnamefont
  {Arganda-Carreras}}, \bibinfo {author} {\bibfnamefont {E.}~\bibnamefont
  {Frise}}, \bibinfo {author} {\bibfnamefont {V.}~\bibnamefont {Kaynig}},
  \bibinfo {author} {\bibfnamefont {M.}~\bibnamefont {Longair}}, \bibinfo
  {author} {\bibfnamefont {T.}~\bibnamefont {Pietzsch}}, \bibinfo {author}
  {\bibfnamefont {S.}~\bibnamefont {Preibisch}}, \bibinfo {author}
  {\bibfnamefont {C.}~\bibnamefont {Rueden}}, \bibinfo {author} {\bibfnamefont
  {S.}~\bibnamefont {Saalfeld}}, \bibinfo {author} {\bibfnamefont
  {B.}~\bibnamefont {Schmid}}, \bibinfo {author} {\bibfnamefont {J.-Y.}\
  \bibnamefont {Tinevez}}, \bibinfo {author} {\bibfnamefont {D.~J.}\
  \bibnamefont {White}}, \bibinfo {author} {\bibfnamefont {V.}~\bibnamefont
  {Hartenstein}}, \bibinfo {author} {\bibfnamefont {K.}~\bibnamefont
  {Eliceiri}}, \bibinfo {author} {\bibfnamefont {P.}~\bibnamefont {Tomancak}},
  \ and\ \bibinfo {author} {\bibfnamefont {A.}~\bibnamefont {Cardona}},\ }\href
  {\doibase 10.1038/nmeth.2019} {\bibfield  {journal} {\bibinfo  {journal}
  {Nature Methods}\ }\textbf {\bibinfo {volume} {9}},\ \bibinfo {pages} {676}
  (\bibinfo {year} {2012})}\BibitemShut {NoStop}%
\bibitem [{\citenamefont {Bayer}(1961)}]{bayer_shallow-water_1961}%
  \BibitemOpen
  \bibfield  {author} {\bibinfo {author} {\bibfnamefont {F.~M.}\ \bibnamefont
  {Bayer}},\ }\href@noop {} {\bibfield  {journal} {\bibinfo  {journal} {Studies
  on the Fauna of Curaçao and other Caribbean Islands}\ }\textbf {\bibinfo
  {volume} {12}},\ \bibinfo {pages} {1} (\bibinfo {year} {1961})}\BibitemShut
  {NoStop}%
\bibitem [{\citenamefont {Cairns}(1977)}]{cairns_checklist_1977}%
  \BibitemOpen
  \bibfield  {author} {\bibinfo {author} {\bibfnamefont {S.~D.}\ \bibnamefont
  {Cairns}},\ }\href@noop {} {\bibfield  {journal} {\bibinfo  {journal} {Gulf
  and Caribbean Research}\ }\textbf {\bibinfo {volume} {6}},\ \bibinfo {pages}
  {9} (\bibinfo {year} {1977})}\BibitemShut {NoStop}%
\bibitem [{\citenamefont {{RodiCS}}(2020)}]{mou3adb_rodics_2020}%
  \BibitemOpen
  \bibfield  {author} {\bibinfo {author} {\bibnamefont {{RodiCS}}},\ }\href
  {\doibase 10.5281/zenodo.4023287} {\enquote {\bibinfo {title} {{RodiCS}: a
  finite element solver of kirchhoff rods under fluid flow and more},}\ }
  (\bibinfo {year} {2020}),\ \bibinfo {note}
  {https://zenodo.org/record/4023287}\BibitemShut {NoStop}%
\bibitem [{\citenamefont {Alnæs}\ \emph {et~al.}(2015)\citenamefont {Alnæs},
  \citenamefont {Blechta}, \citenamefont {Hake}, \citenamefont {Johansson},
  \citenamefont {Kehlet}, \citenamefont {Logg}, \citenamefont {Richardson},
  \citenamefont {Ring}, \citenamefont {Rognes},\ and\ \citenamefont
  {Wells}}]{alnaes_fenics_2015}%
  \BibitemOpen
  \bibfield  {author} {\bibinfo {author} {\bibfnamefont {M.}~\bibnamefont
  {Alnæs}}, \bibinfo {author} {\bibfnamefont {J.}~\bibnamefont {Blechta}},
  \bibinfo {author} {\bibfnamefont {J.}~\bibnamefont {Hake}}, \bibinfo {author}
  {\bibfnamefont {A.}~\bibnamefont {Johansson}}, \bibinfo {author}
  {\bibfnamefont {B.}~\bibnamefont {Kehlet}}, \bibinfo {author} {\bibfnamefont
  {A.}~\bibnamefont {Logg}}, \bibinfo {author} {\bibfnamefont {C.}~\bibnamefont
  {Richardson}}, \bibinfo {author} {\bibfnamefont {J.}~\bibnamefont {Ring}},
  \bibinfo {author} {\bibfnamefont {M.~E.}\ \bibnamefont {Rognes}}, \ and\
  \bibinfo {author} {\bibfnamefont {G.~N.}\ \bibnamefont {Wells}},\ }\href
  {\doibase 10.11588/ans.2015.100.20553} {\bibfield  {journal} {\bibinfo
  {journal} {Archive of Numerical Software}\ }\textbf {\bibinfo {volume} {3}}
  (\bibinfo {year} {2015}),\ 10.11588/ans.2015.100.20553},\ \bibinfo {note}
  {number: 100}\BibitemShut {NoStop}%
\bibitem [{\citenamefont {Taylor}(1952)}]{taylor_analysis_1952}%
  \BibitemOpen
  \bibfield  {author} {\bibinfo {author} {\bibfnamefont {G.~I.}\ \bibnamefont
  {Taylor}},\ }\href {\doibase 10.1098/rspa.1952.0159} {\bibfield  {journal}
  {\bibinfo  {journal} {Proceedings of the Royal Society of London. Series A.
  Mathematical and Physical Sciences}\ }\textbf {\bibinfo {volume} {214}},\
  \bibinfo {pages} {158} (\bibinfo {year} {1952})},\ \bibinfo {note}
  {publisher: Royal Society}\BibitemShut {NoStop}%
\bibitem [{\citenamefont {Facchinetti}\ \emph
  {et~al.}(2004{\natexlab{b}})\citenamefont {Facchinetti}, \citenamefont {{de
  Langre}},\ and\ \citenamefont {Biolley}}]{facchinetti_coupling_2004}%
  \BibitemOpen
  \bibfield  {author} {\bibinfo {author} {\bibfnamefont {M.~L.}\ \bibnamefont
  {Facchinetti}}, \bibinfo {author} {\bibfnamefont {E.}~\bibnamefont {{de
  Langre}}}, \ and\ \bibinfo {author} {\bibfnamefont {F.}~\bibnamefont
  {Biolley}},\ }\href {\doibase 10.1016/j.jfluidstructs.2003.12.004} {\bibfield
   {journal} {\bibinfo  {journal} {Journal of Fluids and Structures}\ }\textbf
  {\bibinfo {volume} {19}},\ \bibinfo {pages} {123} (\bibinfo {year}
  {2004}{\natexlab{b}})}\BibitemShut {NoStop}%
\bibitem [{\citenamefont
  {{NDBC}}(2020)}]{national_data_buoy_center_national_nodate}%
  \BibitemOpen
  \bibfield  {author} {\bibinfo {author} {\bibnamefont {{NDBC}}},\ }\href
  {https://www.ndbc.noaa.gov/} {{\enquote {\bibinfo
  {title} {{https://www.ndbc.noaa.gov/}},}\ }} (\bibinfo {year}
  {2020})\BibitemShut {NoStop}%
\bibitem [{\citenamefont {Blevins}(1990)}]{blevins_flow-induced_1990}%
  \BibitemOpen
  \bibfield  {author} {\bibinfo {author} {\bibfnamefont {R.}~\bibnamefont
  {Blevins}},\ }\href@noop {} {\emph {\bibinfo {title} {Flow-induced
  vibration}}}\ (\bibinfo  {publisher} {Van Nostrand Reinhold},\ \bibinfo
  {year} {1990})\BibitemShut {NoStop}%
\bibitem [{\citenamefont {Huera-Huarte}\ and\ \citenamefont
  {Gharib}(2011)}]{huera-huarte_flow-induced_2011}%
  \BibitemOpen
  \bibfield  {author} {\bibinfo {author} {\bibfnamefont {F.~J.}\ \bibnamefont
  {Huera-Huarte}}\ and\ \bibinfo {author} {\bibfnamefont {M.}~\bibnamefont
  {Gharib}},\ }\href {\doibase 10.1016/j.jfluidstructs.2011.01.001} {\bibfield
  {journal} {\bibinfo  {journal} {Journal of Fluids and Structures}\ }\textbf
  {\bibinfo {volume} {27}},\ \bibinfo {pages} {354} (\bibinfo {year}
  {2011})}\BibitemShut {NoStop}%
\bibitem [{\citenamefont {{Nova South Eastern
  University}}(2016)}]{nova_south_eastern_university_south_2016}%
  \BibitemOpen
  \bibfield  {author} {\bibinfo {author} {\bibnamefont {{Nova South Eastern
  University}}},\ }\href {http://nsuworks.nova.edu/octocoral_guide/}
  {{\enquote {\bibinfo {title} {South {Florida}
  {Octocorals}: {A} {Guide} to {Identification}},}\ }} (\bibinfo {year}
  {2016})\BibitemShut {NoStop}%
\bibitem [{\citenamefont {Nakamura}\ \emph {et~al.}(1994)\citenamefont
  {Nakamura}, \citenamefont {Hirata},\ and\ \citenamefont
  {Kashima}}]{nakamura_galloping_1994}%
  \BibitemOpen
  \bibfield  {author} {\bibinfo {author} {\bibfnamefont {Y.}~\bibnamefont
  {Nakamura}}, \bibinfo {author} {\bibfnamefont {K.}~\bibnamefont {Hirata}}, \
  and\ \bibinfo {author} {\bibfnamefont {K.}~\bibnamefont {Kashima}},\ }\href
  {\doibase 10.1006/jfls.1994.1017} {\bibfield  {journal} {\bibinfo  {journal}
  {Journal of Fluids and Structures}\ }\textbf {\bibinfo {volume} {8}},\
  \bibinfo {pages} {355} (\bibinfo {year} {1994})}\BibitemShut {NoStop}%
\bibitem [{\citenamefont {Pa\"{\i}doussis}\ \emph {et~al.}(2010)\citenamefont
  {Pa\"{\i}doussis}, \citenamefont {Price},\ and\ \citenamefont {{de
  }Langre}}]{paidoussis_fluid-structure_2010}%
  \BibitemOpen
  \bibfield  {author} {\bibinfo {author} {\bibfnamefont {M.}~\bibnamefont
  {Pa\"{\i}doussis}}, \bibinfo {author} {\bibfnamefont {S.}~\bibnamefont
  {Price}}, \ and\ \bibinfo {author} {\bibfnamefont {E.}~\bibnamefont {{de
  }Langre}},\ }\href@noop {} {\emph {\bibinfo {title} {Fluid-{Structure}
  {Interactions}: {Cross}-{Flow}-{Induced} {Instabilities}}}}\ (\bibinfo
  {publisher} {Cambridge University Press},\ \bibinfo {year}
  {2010})\BibitemShut {NoStop}%
\bibitem [{\citenamefont {Etienne}\ \emph {et~al.}(2009)\citenamefont
  {Etienne}, \citenamefont {Garon},\ and\ \citenamefont
  {Pelletier}}]{etienne_perspective_2009}%
  \BibitemOpen
  \bibfield  {author} {\bibinfo {author} {\bibfnamefont {S.}~\bibnamefont
  {Etienne}}, \bibinfo {author} {\bibfnamefont {A.}~\bibnamefont {Garon}}, \
  and\ \bibinfo {author} {\bibfnamefont {D.}~\bibnamefont {Pelletier}},\ }\href
  {\doibase 10.1016/j.jcp.2008.11.032} {\bibfield  {journal} {\bibinfo
  {journal} {Journal of Computational Physics}\ }\textbf {\bibinfo {volume}
  {228}},\ \bibinfo {pages} {2313} (\bibinfo {year} {2009})}\BibitemShut
  {NoStop}%
\bibitem [{\citenamefont {Persillon}\ and\ \citenamefont
  {Braza}(1998)}]{persillon_physical_1998}%
  \BibitemOpen
  \bibfield  {author} {\bibinfo {author} {\bibfnamefont {H.}~\bibnamefont
  {Persillon}}\ and\ \bibinfo {author} {\bibfnamefont {M.}~\bibnamefont
  {Braza}},\ }\href {\doibase 10.1017/S0022112098001116} {\bibfield  {journal}
  {\bibinfo  {journal} {Journal of Fluid Mechanics}\ }\textbf {\bibinfo
  {volume} {365}},\ \bibinfo {pages} {23} (\bibinfo {year} {1998})}\BibitemShut
  {NoStop}%
\bibitem [{\citenamefont {Taylor}\ and\ \citenamefont
  {Hood}(1973)}]{taylor_numerical_1973}%
  \BibitemOpen
  \bibfield  {author} {\bibinfo {author} {\bibfnamefont {C.}~\bibnamefont
  {Taylor}}\ and\ \bibinfo {author} {\bibfnamefont {P.}~\bibnamefont {Hood}},\
  }\href {\doibase 10.1016/0045-7930(73)90027-3} {\bibfield  {journal}
  {\bibinfo  {journal} {Computers \& Fluids}\ }\textbf {\bibinfo {volume}
  {1}},\ \bibinfo {pages} {73} (\bibinfo {year} {1973})}\BibitemShut {NoStop}%
\bibitem [{\citenamefont {Maxey}\ and\ \citenamefont
  {Riley}(1983)}]{maxey1983}%
  \BibitemOpen
  \bibfield  {author} {\bibinfo {author} {\bibfnamefont {M.~R.}\ \bibnamefont
  {Maxey}}\ and\ \bibinfo {author} {\bibfnamefont {J.~J.}\ \bibnamefont
  {Riley}},\ }\href {\doibase 10.1063/1.864230} {\bibfield  {journal} {\bibinfo
   {journal} {The Physics of Fluids}\ }\textbf {\bibinfo {volume} {26}},\
  \bibinfo {pages} {883} (\bibinfo {year} {1983})}\BibitemShut {NoStop}%
\bibitem [{\citenamefont {Ounis}\ and\ \citenamefont
  {Ahmadi}(1990)}]{ounis1990}%
  \BibitemOpen
  \bibfield  {author} {\bibinfo {author} {\bibfnamefont {H.}~\bibnamefont
  {Ounis}}\ and\ \bibinfo {author} {\bibfnamefont {G.}~\bibnamefont {Ahmadi}},\
  }\href {\doibase 10.1115/1.2909358} {\bibfield  {journal} {\bibinfo
  {journal} {Journal of Fluids Engineering}\ }\textbf {\bibinfo {volume}
  {112}},\ \bibinfo {pages} {114} (\bibinfo {year} {1990})}\BibitemShut
  {NoStop}%
\bibitem [{\citenamefont {Barton}(1995)}]{barton1995}%
  \BibitemOpen
  \bibfield  {author} {\bibinfo {author} {\bibfnamefont {I.~E.}\ \bibnamefont
  {Barton}},\ }\href {\doibase 10.1016/0021-8502(95)00018-8} {\bibfield
  {journal} {\bibinfo  {journal} {Journal of Aerosol Science}\ }\textbf
  {\bibinfo {volume} {26}},\ \bibinfo {pages} {887} (\bibinfo {year}
  {1995})}\BibitemShut {NoStop}%
\bibitem [{\citenamefont {Clift}\ \emph {et~al.}(1978)\citenamefont {Clift},
  \citenamefont {Grace},\ and\ \citenamefont {Weber}}]{clift_bubbles_1978}%
  \BibitemOpen
  \bibfield  {author} {\bibinfo {author} {\bibfnamefont {R.}~\bibnamefont
  {Clift}}, \bibinfo {author} {\bibfnamefont {J.~R.}\ \bibnamefont {Grace}}, \
  and\ \bibinfo {author} {\bibfnamefont {M.~E.}\ \bibnamefont {Weber}},\
  }\href@noop {} {\bibfield  {journal} {\bibinfo  {journal} {New York}\
  }\textbf {\bibinfo {volume} {510}},\ \bibinfo {pages} {147} (\bibinfo {year}
  {1978})}\BibitemShut {NoStop}%
\bibitem [{\citenamefont {Brennen}(1982)}]{brennen_review_1982}%
  \BibitemOpen
  \bibfield  {author} {\bibinfo {author} {\bibfnamefont {C.~E.}\ \bibnamefont
  {Brennen}},\ }\href {https://apps.dtic.mil/docs/citations/ADA110190}
  {{\emph {\bibinfo {title} {A {Review} of {Added} {Mass}
  and {Fluid} {Inertial} {Forces}.}}}},\ \bibinfo {type} {Tech. Rep.}\
  (\bibinfo  {institution} {Naval Civil Engineering Laboratory, Port Hueneme,
  California},\ \bibinfo {year} {1982})\BibitemShut {NoStop}%
\bibitem [{\citenamefont {B\'{e}guin}\ \emph {et~al.}(2016)\citenamefont
  {B\'{e}guin}, \citenamefont {Pelletier},\ and\ \citenamefont
  {\'{E}tienne}}]{beguin_void_2016}%
  \BibitemOpen
  \bibfield  {author} {\bibinfo {author} {\bibfnamefont {C.}~\bibnamefont
  {B\'{e}guin}}, \bibinfo {author} {\bibfnamefont {E.}~\bibnamefont
  {Pelletier}}, \ and\ \bibinfo {author} {\bibfnamefont {S.}~\bibnamefont
  {\'{E}tienne}},\ }\href {\doibase 10.1016/j.euromechflu.2015.11.008}
  {\bibfield  {journal} {\bibinfo  {journal} {European Journal of Mechanics -
  B/Fluids}\ }\textbf {\bibinfo {volume} {56}},\ \bibinfo {pages} {28}
  (\bibinfo {year} {2016})}\BibitemShut {NoStop}%
\bibitem [{\citenamefont {Espinosa-Gayosso}\ \emph {et~al.}(2015)\citenamefont
  {Espinosa-Gayosso}, \citenamefont {Ghisalberti}, \citenamefont {Ivey},\ and\
  \citenamefont {Jones}}]{espinosa-gayosso_density-ratio_2015}%
  \BibitemOpen
  \bibfield  {author} {\bibinfo {author} {\bibfnamefont {A.}~\bibnamefont
  {Espinosa-Gayosso}}, \bibinfo {author} {\bibfnamefont {M.}~\bibnamefont
  {Ghisalberti}}, \bibinfo {author} {\bibfnamefont {G.~N.}\ \bibnamefont
  {Ivey}}, \ and\ \bibinfo {author} {\bibfnamefont {N.~L.}\ \bibnamefont
  {Jones}},\ }\href {\doibase 10.1017/jfm.2015.557} {\bibfield  {journal}
  {\bibinfo  {journal} {Journal of Fluid Mechanics}\ }\textbf {\bibinfo
  {volume} {783}},\ \bibinfo {pages} {191} (\bibinfo {year}
  {2015})}\BibitemShut {NoStop}%
\bibitem [{\citenamefont {Hay}\ \emph {et~al.}(2014)\citenamefont {Hay},
  \citenamefont {Yu}, \citenamefont {Etienne}, \citenamefont {Garon},\ and\
  \citenamefont {Pelletier}}]{hay2014}%
  \BibitemOpen
  \bibfield  {author} {\bibinfo {author} {\bibfnamefont {A.}~\bibnamefont
  {Hay}}, \bibinfo {author} {\bibfnamefont {K.~R.}\ \bibnamefont {Yu}},
  \bibinfo {author} {\bibfnamefont {S.}~\bibnamefont {Etienne}}, \bibinfo
  {author} {\bibfnamefont {A.}~\bibnamefont {Garon}}, \ and\ \bibinfo {author}
  {\bibfnamefont {D.}~\bibnamefont {Pelletier}},\ }\href {\doibase
  10.1016/j.compfluid.2014.04.036} {\bibfield  {journal} {\bibinfo  {journal}
  {Computers \& Fluids}\ }\textbf {\bibinfo {volume} {100}},\ \bibinfo {pages}
  {204} (\bibinfo {year} {2014})}\BibitemShut {NoStop}%
\bibitem [{\citenamefont {Yu}\ \emph {et~al.}(2015)\citenamefont {Yu},
  \citenamefont {{\'E}tienne}, \citenamefont {Hay},\ and\ \citenamefont
  {Pelletier}}]{yu2015}%
  \BibitemOpen
  \bibfield  {author} {\bibinfo {author} {\bibfnamefont {K.~R.}\ \bibnamefont
  {Yu}}, \bibinfo {author} {\bibfnamefont {S.}~\bibnamefont {{\'E}tienne}},
  \bibinfo {author} {\bibfnamefont {A.}~\bibnamefont {Hay}}, \ and\ \bibinfo
  {author} {\bibfnamefont {D.}~\bibnamefont {Pelletier}},\ }\href {\doibase
  10.1007/s00162-015-0367-4} {\bibfield  {journal} {\bibinfo  {journal}
  {Theoretical and Computational Fluid Dynamics}\ }\textbf {\bibinfo {volume}
  {29}},\ \bibinfo {pages} {455} (\bibinfo {year} {2015})}\BibitemShut
  {NoStop}%
\bibitem [{\citenamefont {Cori}\ \emph {et~al.}(2015)\citenamefont {Cori},
  \citenamefont {Etienne}, \citenamefont {Garon},\ and\ \citenamefont
  {Pelletier}}]{cori2015}%
  \BibitemOpen
  \bibfield  {author} {\bibinfo {author} {\bibfnamefont {J.-F.}\ \bibnamefont
  {Cori}}, \bibinfo {author} {\bibfnamefont {S.}~\bibnamefont {Etienne}},
  \bibinfo {author} {\bibfnamefont {A.}~\bibnamefont {Garon}}, \ and\ \bibinfo
  {author} {\bibfnamefont {D.}~\bibnamefont {Pelletier}},\ }\href {\doibase
  10.1002/fld.4020} {\bibfield  {journal} {\bibinfo  {journal} {International
  Journal for Numerical Methods in Fluids}\ }\textbf {\bibinfo {volume} {78}},\
  \bibinfo {pages} {385} (\bibinfo {year} {2015})}\BibitemShut {NoStop}%
\bibitem [{\citenamefont {Hay}\ \emph {et~al.}(2015{\natexlab{a}})\citenamefont
  {Hay}, \citenamefont {Etienne}, \citenamefont {Garon},\ and\ \citenamefont
  {Pelletier}}]{hay2015}%
  \BibitemOpen
  \bibfield  {author} {\bibinfo {author} {\bibfnamefont {A.}~\bibnamefont
  {Hay}}, \bibinfo {author} {\bibfnamefont {S.}~\bibnamefont {Etienne}},
  \bibinfo {author} {\bibfnamefont {A.}~\bibnamefont {Garon}}, \ and\ \bibinfo
  {author} {\bibfnamefont {D.}~\bibnamefont {Pelletier}},\ }\href {\doibase
  10.1016/j.cma.2015.06.006} {\bibfield  {journal} {\bibinfo  {journal}
  {Computer Methods in Applied Mechanics and Engineering}\ }\textbf {\bibinfo
  {volume} {295}},\ \bibinfo {pages} {172} (\bibinfo {year}
  {2015}{\natexlab{a}})}\BibitemShut {NoStop}%
\bibitem [{\citenamefont {Yu}\ \emph {et~al.}(2016)\citenamefont {Yu},
  \citenamefont {{\'E}tienne}, \citenamefont {Scolan}, \citenamefont {Hay},
  \citenamefont {Fontaine},\ and\ \citenamefont {Pelletier}}]{yu2016}%
  \BibitemOpen
  \bibfield  {author} {\bibinfo {author} {\bibfnamefont {K.~R.}\ \bibnamefont
  {Yu}}, \bibinfo {author} {\bibfnamefont {S.}~\bibnamefont {{\'E}tienne}},
  \bibinfo {author} {\bibfnamefont {Y.-M.}\ \bibnamefont {Scolan}}, \bibinfo
  {author} {\bibfnamefont {A.}~\bibnamefont {Hay}}, \bibinfo {author}
  {\bibfnamefont {E.}~\bibnamefont {Fontaine}}, \ and\ \bibinfo {author}
  {\bibfnamefont {D.}~\bibnamefont {Pelletier}},\ }\href {\doibase
  10.1016/j.jfluidstructs.2015.10.005} {\bibfield  {journal} {\bibinfo
  {journal} {Journal of Fluids and Structures}\ }\textbf {\bibinfo {volume}
  {60}},\ \bibinfo {pages} {37} (\bibinfo {year} {2016})}\BibitemShut {NoStop}%
\bibitem [{\citenamefont {Hay}\ \emph {et~al.}(2015{\natexlab{b}})\citenamefont
  {Hay}, \citenamefont {Etienne}, \citenamefont {Pelletier},\ and\
  \citenamefont {Garon}}]{hay_hp-adaptive_2015}%
  \BibitemOpen
  \bibfield  {author} {\bibinfo {author} {\bibfnamefont {A.}~\bibnamefont
  {Hay}}, \bibinfo {author} {\bibfnamefont {S.}~\bibnamefont {Etienne}},
  \bibinfo {author} {\bibfnamefont {D.}~\bibnamefont {Pelletier}}, \ and\
  \bibinfo {author} {\bibfnamefont {A.}~\bibnamefont {Garon}},\ }\href
  {\doibase 10.1016/j.jcp.2015.03.022} {\bibfield  {journal} {\bibinfo
  {journal} {Journal of Computational Physics}\ }\textbf {\bibinfo {volume}
  {291}},\ \bibinfo {pages} {151} (\bibinfo {year}
  {2015}{\natexlab{b}})}\BibitemShut {NoStop}%
\bibitem [{\citenamefont
  {{\textsc{Paradvect}}}(2020)}]{mou3adb_paradvect_2020}%
  \BibitemOpen
  \bibfield  {author} {\bibinfo {author} {\bibnamefont
  {{\textsc{Paradvect}}}},\ }\href {\doibase 10.5281/zenodo.3981610} {\enquote
  {\bibinfo {title} {{\textsc{Paradvect} (PARticle ADVECTion): a Python code to
  simulate the trajectory of particles advected by a fluid flow}},}\ }
  (\bibinfo {year} {2020})\BibitemShut {NoStop}%
\bibitem [{\citenamefont {Löhner}\ and\ \citenamefont
  {Ambrosiano}(1990)}]{lohner_vectorized_1990}%
  \BibitemOpen
  \bibfield  {author} {\bibinfo {author} {\bibfnamefont {R.}~\bibnamefont
  {Löhner}}\ and\ \bibinfo {author} {\bibfnamefont {J.}~\bibnamefont
  {Ambrosiano}},\ }\href {\doibase 10.1016/0021-9991(90)90002-I} {\bibfield
  {journal} {\bibinfo  {journal} {Journal of Computational Physics}\ }\textbf
  {\bibinfo {volume} {91}},\ \bibinfo {pages} {22} (\bibinfo {year}
  {1990})}\BibitemShut {NoStop}%
\bibitem [{\citenamefont {Löhner}(2008)}]{lohner_applied_2008}%
  \BibitemOpen
  \bibfield  {author} {\bibinfo {author} {\bibfnamefont {R.}~\bibnamefont
  {L\"{o}hner}},\ }\href@noop {} {{\emph {\bibinfo {title}
  {Applied {Computational} {Fluid} {Dynamics} {Techniques}: {An} {Introduction}
  {Based} on {Finite} {Element} {Methods}}}}}\ (\bibinfo  {publisher} {John
  Wiley \& Sons},\ \bibinfo {year} {2008})\BibitemShut {NoStop}%
\bibitem [{\citenamefont {Buckingham}(1914)}]{buckingham_physically_1914}%
  \BibitemOpen
  \bibfield  {author} {\bibinfo {author} {\bibfnamefont {E.}~\bibnamefont
  {Buckingham}},\ }\href {\doibase 10.1103/PhysRev.4.345} {\bibfield  {journal}
  {\bibinfo  {journal} {Physical Review}\ }\textbf {\bibinfo {volume} {4}},\
  \bibinfo {pages} {345} (\bibinfo {year} {1914})},\ \bibinfo {note}
  {publisher: American Physical Society}\BibitemShut {NoStop}%
\bibitem [{\citenamefont {Shimeta}\ and\ \citenamefont
  {Jumars}(1991)}]{shimeta_physical_1991}%
  \BibitemOpen
  \bibfield  {author} {\bibinfo {author} {\bibfnamefont {J.}~\bibnamefont
  {Shimeta}}\ and\ \bibinfo {author} {\bibfnamefont {P.~A.}\ \bibnamefont
  {Jumars}},\ }\href@noop {} {\bibfield  {journal} {\bibinfo  {journal}
  {Oceanogr. Mar. Biol. Annu. Rev}\ }\textbf {\bibinfo {volume} {29}},\
  \bibinfo {pages} {l} (\bibinfo {year} {1991})}\BibitemShut {NoStop}%
\bibitem [{\citenamefont {Boudina}\ \emph {et~al.}(2020)\citenamefont
  {Boudina}, \citenamefont {Gosselin},\ and\ \citenamefont
  {{\'E}tienne}}]{boudina2020}%
  \BibitemOpen
  \bibfield  {author} {\bibinfo {author} {\bibfnamefont {M.}~\bibnamefont
  {Boudina}}, \bibinfo {author} {\bibfnamefont {F.~P.}\ \bibnamefont
  {Gosselin}}, \ and\ \bibinfo {author} {\bibfnamefont {S.}~\bibnamefont
  {{\'E}tienne}},\ }\href {\doibase 10.1063/5.0030891} {\bibfield  {journal}
  {\bibinfo  {journal} {Physics of Fluids}\ }\textbf {\bibinfo {volume} {32}},\
  \bibinfo {pages} {123603} (\bibinfo {year} {2020})}\BibitemShut {NoStop}%
\bibitem [{\citenamefont {Khalak}\ and\ \citenamefont
  {Williamson}(1999)}]{khalak_motions_1999}%
  \BibitemOpen
  \bibfield  {author} {\bibinfo {author} {\bibfnamefont {A.}~\bibnamefont
  {Khalak}}\ and\ \bibinfo {author} {\bibfnamefont {C.~H.~K.}\ \bibnamefont
  {Williamson}},\ }\href {\doibase 10.1006/jfls.1999.0236} {\bibfield
  {journal} {\bibinfo  {journal} {Journal of Fluids and Structures}\ }\textbf
  {\bibinfo {volume} {13}},\ \bibinfo {pages} {813} (\bibinfo {year}
  {1999})}\BibitemShut {NoStop}%
\end{thebibliography}%
